\documentclass[
	aps,
	pre,
	reprint,
	10pt,
	floatfix,
	superscriptaddress,
	longbibliography
]{revtex4-2}

\newcommand{\myTitle}{Free-Energy Transduction in Chemical Reaction Networks: from Enzymes to Metabolism}

\usepackage[utf8]{inputenc}

\usepackage[lining,semibold]{libertine} 
\usepackage[libertine, cmintegrals, bigdelims, vvarbb]{newtxmath}
\usepackage{booktabs}

\usepackage{calc}
\usepackage{bm}
\usepackage{braket}
\usepackage{microtype}
\usepackage{iftex}
\usepackage{hyperref}
\usepackage{xcolor}
\usepackage{graphicx}
\usepackage{siunitx}

\usepackage[version=4]{mhchem}

\usepackage[normalem]{ulem}

\definecolor{webgreen}{rgb}{0,.5,0}
\definecolor{webbrown}{rgb}{.6,0,0}
\definecolor{grigio}{rgb}{.85,.85,.85} 
\definecolor{RoyalBlue}{rgb}{0.0, 0.14, 0.4}

\definecolor{butter1}{rgb}{0.98,0.91,0.31}
\definecolor{butter2}{rgb}{0.93,0.83,0}
\definecolor{butter3}{rgb}{0.77,0.63,0}
\definecolor{skyblue1}{rgb}{0.45,0.62,0.81}
\definecolor{skyblue2}{rgb}{0.2,0.39,0.64}
\definecolor{skyblue3}{rgb}{0.13,0.29,0.53}
\definecolor{scarlet1}{rgb}{0.93,0.16,0.16}
\definecolor{scarlet2}{rgb}{0.8,0,0}
\definecolor{scarlet3}{rgb}{0.64,0,0}
\definecolor{chameleon1}{rgb}{0.54,0.88,0.2}
\definecolor{chameleon2}{rgb}{0.45,0.82,0.09}
\definecolor{chameleon3}{rgb}{0.3,0.6,0.02}
\definecolor{orange1}{rgb}{0.98,0.68,0.24}
\definecolor{orange2}{rgb}{0.96,0.47,0}
\definecolor{orange3}{rgb}{0.8,0.36,0}
\definecolor{plum1}{rgb}{0.68,0.5,0.66}
\definecolor{plum2}{rgb}{0.46,0.31,0.48}
\definecolor{plum3}{rgb}{0.36,0.21,0.4}
\definecolor{chocolate1}{rgb}{0.91,0.72,0.43}
\definecolor{chocolate2}{rgb}{0.75,0.49,0.07}
\definecolor{chocolate3}{rgb}{0.56,0.35,0.01}
\definecolor{aluminium1}{rgb}{0.93,0.93,0.92}
\definecolor{aluminium2}{rgb}{0.82,0.84,0.81}
\definecolor{aluminium3}{rgb}{0.73,0.74,0.71}
\definecolor{aluminium4}{rgb}{0.53,0.54,0.52}
\definecolor{aluminium5}{rgb}{0.33,0.34,0.32}
\definecolor{aluminium6}{rgb}{0.18,0.2,0.21}  

\hypersetup{%
    colorlinks=true, linktocpage=true, pdfstartpage=1, pdfstartview=FitV,%
    breaklinks=true, pdfpagemode=UseNone, pageanchor=true, pdfpagemode=UseOutlines,%
    plainpages=false, bookmarksnumbered, bookmarksopen=true, bookmarksopenlevel=1,%
    hypertexnames=true, pdfhighlight=/O,
    urlcolor=webbrown, linkcolor=RoyalBlue, citecolor=webgreen, 
    pdftitle={\myTitle},%
    pdfauthor={\textcopyright\ The Authors},%
    pdfsubject={},%
    pdfkeywords={},%
    pdfcreator={pdfLaTeX}%
}

\ifPDFTeX
  
\else
  
\fi

\newcommand{\stoich}{\ensuremath{\mathbb{S}}}

\newcommand{\updownharpoons}{\rotatebox[origin=c]{-90}{$\rightleftharpoons$}}

\renewcommand{\vec}[1]{\bm{#1}}

\newcommand{\subsc}[1]{_{\text{#1}}}
\newcommand{\supsc}[1]{^{\text{#1}}}
\newcommand{\T}{^{\mathsf{T}}}

\newcommand{\eq}[1]{(\ref{eq:#1})}
\newcommand{\Eq}[1]{Eq.~\eq{#1}}
\newcommand{\Eqs}[1]{Eqs.~\eq{#1}}

\newcommand{\Sec}[1]{Sec.~\ref{sec:#1}}

\newcommand{\App}[1]{App.~\ref{sec:#1}}

\newcommand{\pf}{p_{1}}
\newcommand{\ps}{p_{2}}
\newcommand{\ef}{\varepsilon_{1}}
\newcommand{\es}{\varepsilon_{2}}

\begin{document}

\date{\today}

\title{\myTitle}

\author{Artur Wachtel}
\affiliation{Department of Molecular, Cellular and Developmental Biology, Yale University, New Haven (CT), U.S.A.}
\author{Riccardo Rao}
\affiliation{Simons Center for Systems Biology, School of Natural Sciences, Institute for Advanced Study, 08540 Princeton (NJ), U.S.A.}
\author{Massimiliano Esposito}
\email{massimiliano.esposito@uni.lu}
\affiliation{
Complex Systems and Statistical Mechanics, Department of Physics and Materials Science, University of Luxembourg, 162a, Avenue de la Fa\"{i}encerie, 1511 Luxembourg, G. D. Luxembourg
}

\begin{abstract}
We provide a rigorous definition of free-energy transduction and its efficiency in arbitrary---linear or nonlinear---open chemical reaction networks (CRNs) operating at steady state.
Our method is based on the knowledge of the stoichiometric matrix and of the chemostatted species (\textit{i.e.} the species maintained at constant concentration by the environment) to identify the fundamental currents and forces contributing to the entropy production.
Transduction occurs when the current of a stoichiometrically balanced process is driven against its spontaneous direction (set by its force) thanks to other processes flowing along their spontaneous direction.
In these regimes, open CRNs operate as thermodynamic machines.
After exemplifying these general ideas using toy models, we analyze central energy metabolism.
We relate the fundamental currents to metabolic pathways and discuss the efficiency with which they are able to transduce free energy.
\end{abstract}

\maketitle

\section{Introduction}
\label{sec:intro}

Living organisms are out-of-equilibrium systems which harness chemical free energy from their surrounding to fuel their own function \cite{yang21}.
To do so, free energy from external energy-rich molecules is transferred into energetic mediator molecules---primarily adenosine triphosphate (ATP)---which are then used to power the otherwise-unfavorable reactions that are essential for life.
The resulting low-energy molecules and heat are released back into the environment as waste.
At the core of this process lies energy metabolism:
a finely-regulated network of coupled biochemical reactions that---through enzymatic catalysis---enable transformations on the environment that would hardly occur spontaneously \cite{nelson08,voet16}.

In order to study the thermodynamics of living systems, traditional chemical thermodynamics is of limited use.
Indeed it considers transformations between initial and final equilibrium distributions of chemicals (typically the equilibrium state before and after reactant and product are brought in contact) and time is absent from the description.
In contrast, energy metabolism is a dynamical process that never reaches equilibrium, except upon death of the living organism.
The thermodynamics of living systems is thus much more akin to that of thermodynamic machines, which transduce free energy thanks to reservoirs at different temperatures, pressures or chemical potentials.
These chemical machines must be studied using modern formulations of nonequilibrium thermodynamics, as these formulations combine phenomenological irreversible thermodynamics with dynamical rate laws. 
To analyze their performance, identifying the conjugated pairs of thermodynamic force and current which determine the entropy production (or dissipation) is a crucial step.
While this is easily done for macroscopic machines \cite{bejan16}, as well as for simple biological or synthetic molecular machines \cite{Hill1977,Wachtel.etal2018,Juelicher1997,Parmeggiani1999,Parrondo2002,Seifert2012,Brown2020,amano21}, slightly more complex machines described by stochastic dynamics on finite state graphs already require some graph theoretical methods, like those pioneered by T.L.~Hill in his work on free-energy transduction~\cite{Hill1977} as well as by J.~Schnakenberg \cite{schnakenberg76,andrieux07:schnakenberg}.
Furthermore, for chemical reaction networks (CRNs), these theories are restricted to chemical systems with elementary first or pseudo-first order reactions, \textit{i.e.} systems whose deterministic dynamics in terms of concentrations is linear and equivalent to the dynamics for the probability to find a molecule of a given type.
These restrictions can be justified when studying the energetics of single enzymes, but become completely inadequate to describe even simple metabolic networks, where many reactions contain multimolecular steps and are described by enzymatic rate laws (\textit{i.e.} for non-elementary reactions).

In this paper, we generalize these ideas, leaving these limitations behind. 
We provide a general theory of free-energy transduction for any open CRN.
To do so, we make use of the topology of the networks (encoded in the stoichiometric matrix) and of the chemostatted species to determine the fundamental thermodynamic forces and currents that are necessary to define free-energy transduction and its efficiency.
We then demonstrate the value of the theory by analyzing free-energy transduction in central metabolism, highlighting the relation between metabolic pathways and fundamental thermodynamic currents.

In section~\ref{sec:chem-networks} we review the dynamics and nonequilibrium thermodynamics of open CRNs. In particular, we show how the stoichiometric matrix can be used to identify the fundamental thermodynamic forces and currents controlling the dissipation. 
In section~\ref{sec:coupling} we show how to make use of these results to analyze simple and illustrative CRNs, thus paving the way for a systematic study of free-energy transduction and its efficiency.
In section~\ref{sec:metabolism} we show that our method is not limited to the analysis of simple systems by applying it to central energy metabolism.
We first consider glycolysis, the tricarboxylic acid cycle and the electron transport chain, separately and combined.
Ethanol fermentation and gluconeogenesis are subsequently discussed.
We highlight the relation between metabolic pathways and fundamental currents and analyze their efficiency.
In section~\ref{sec:transduction} we formalize the systematic procedure to analyze free-energy transduction and its efficiency in arbitrary CRNs.
Section~\ref{Conc} contains a discussion and the conclusions.


\section{Chemical Reaction Networks}
\label{sec:chem-networks}

After defining the stoichiometry and kinetics, we review how to use thermodynamics to analyze the dissipation caused by exchanges of free energy with the environment \cite{Polettini.Esposito2014,Rao.Esposito2016,Wachtel.etal2018,avanzini20,avanzini21}.

\subsection{Stoichiometry}

A CRN describes the interconversion of a set of chemical species \(\ce{Z1}\), \dots, \(\ce{Z_n}\) by means of a set of chemical reactions \(\rho_1, \dots, \rho_m\).
Each reaction \(\rho_j\) is characterized by an equation
\begin{align}
  \nu_{1, j} \ce{Z_1}
  + \dots
  + \nu_{n, j} \ce{Z_n}
  &\rightleftharpoons
  \bar{\nu}_{1, j} \ce{Z_1}
  + \dots
  + \bar{\nu}_{n, j} \ce{Z_n} \, ,
  \label{eq:crn:stoichiometry}
\end{align}
which encodes that each time reaction \(\rho_j\) occurs, it will destroy \(\nu_{i, j} \geq 0\) reactant molecules of type \(\ce{Z_i}\) and produce \(\bar{\nu}_{k, j} \geq 0\) product molecules of type \(\ce{Z_k}\).
Their differences \(\stoich_{i, j} = \bar{\nu}_{i, j} - \nu_{i, j}\) are the \emph{stoichiometric coefficients} that represent the net production of each species in each reaction.
We combine the latter into an \(n \times m\) matrix \(\stoich\) called \emph{stoichiometric matrix}.

As imposed by the conservation laws of physical chemistry, we always assume that any such reaction is balanced in terms of electrons and atomic nuclei, and consequently also balanced in terms of mass and charge: every proton needs to be explicitly included in the stoichiometry.
Furthermore every reaction is reversible: each reaction \(\rho_{+j}\) has a naturally associated reverse reaction \(\rho_{-j}\) where reactants and products are interchanged.

\subsection{Dynamics}

In addition to the stoichiometry encoded in the stoichiometric matrix \(\stoich\), each reaction \(\rho_j\) comes with a \emph{reaction current} \(J_j(\vec{z})\).
These currents characterize the net rate of reaction when the species concentrations are \(\vec{z}\).
The precise expression of \(\vec{J}(\vec{z})\) is determined by the reaction \emph{kinetics}, which depends on both the type of reaction (elementary, enzyme-catalyzed, \textit{etc.}), and on the type of solution (ideal dilute solution, non-ideal dilute solution, \textit{etc.}). 

The reaction currents and the stoichiometric coefficients govern the dynamics of the concentrations \(\vec{z}\) via the deterministic differential equations,
\begin{align}
  \dot{\vec{z}} = \stoich \vec{J}(\vec{z}) + \vec{I}(\vec{z}) \,,
\end{align}
known as the \emph{rate equations} of chemical kinetics.
In open CRNs, \(\vec{I}(\vec{z})\) accounts for the external sources and sinks.
\(\vec{I}(\vec{z})\) is absent in closed CRNs.  
Both here and in the following it is sufficient for our purposes to limit our discussion to steady states, where concentrations are stationary: \( \vec{0} = \dot{\vec{z}} = \stoich \vec{J} + \vec{I}\) (the dependence on \(\vec{z}\) is sometimes omitted for brevity).
We will briefly discuss extensions beyond stationarity in the discussion.
In case of stationarity, reaction currents and external currents have to balance out: \(\stoich\vec{J}= - \vec{I}\).
These steady states are in general out of thermodynamic equilibrium.

\subsection{Thermodynamics}

We consider dilute solutions kept at constant temperature \(T\).
For each chemical species \(i\), the chemical potential \(\mu_i(\vec{z})\) characterizes the free-energy content of that species when the concentrations are \(\vec{z}\)~\footnote{For isobaric solutions, \(\mu_{i}(\vec{z})\) must be regarded as the Gibbs free-energy content of species \(i\)}.
As for the kinetics, the precise expression of \(\mu_i(\vec{z})\) depends on the type of solution.
For instance, for ideal dilute solutions, \(\mu_i(\vec{z})\) solely depends on the concentration of $i$,
\begin{align}
	\mu_i(\vec{z}) = \mu^{\circ}_i + RT \ln z_i = RT \ln \frac{z_i}{z_{i,\mathrm{eq}}} \, ,
	\label{eq:crn:chempot}
\end{align}
where $\mu^\circ_{i}$ is the standard-state chemical potential of species \(i\), \(R\) is the gas constant; see \textit{e.g.\@} \cite{Alberty2003}.
For non-ideal dilute solutions \(\mu_i(\vec{z})\) additionally depends on the concentrations of all chemicals via an activity coefficient \(\gamma_i(\vec{z})\), which effectively accounts for the non-negligible interactions between chemicals,
\begin{equation}
	\mu_{i}(\vec{z}) = \mu^\circ_{i} + RT \ln z_{i} + RT \ln \gamma_{i}(\vec{z}) \, .
	\label{eq:mu}
\end{equation}

The stoichiometric difference of chemical potentials serves as a thermodynamic force for the reactions:
\begin{align}
	-\Delta_j G(\vec{z}) = -\stoich_{j}\T \vec{\mu}(\vec{z}) \, ,
	\label{eq:crn:force}
\end{align}
where \(\stoich_j\) is the \(j\)th colum of the stoichiometric matrix.
The overall dissipation in the reaction network is given by the product of forces and currents for all reactions:
\begin{align}
	T \sigma = -\sum_j  J_j \, \Delta_j G \geq 0 \, .
	\label{eq:crn:dissipation}
\end{align}
The \emph{entropy production rate} (EPR) $\sigma$ quantifies the rate of entropy change in the system and in the reservoirs (i.e. the solution and the chemostats) \cite{Rao.Esposito2016}.
The second law of thermodynamics guarantees that $\sigma$ is non-negative, and that it vanishes solely at thermodynamic equilibrium, where the thermodynamic forces of all reactions vanish. 

In general, the \emph{dissipative contribution} of each reaction, \(- J_j \, \Delta_j G\), need not be non-negative.
As we will discuss soon, the presence of negative contributions is indicative of transduction mechanisms.
At the same time, there are important types of reaction for which \(- J_j \, \Delta_j G\) is always non-negative:
elementary reactions in ideal and non-ideal solutions \cite{Rao.Esposito2016,avanzini21}, as well as cytosolic enzymes \cite{Wachtel.etal2018,avanzini20}.
For these types of reaction, the kinetics and thermodynamics are linked by the \emph{local detailed balance conditions}
\begin{subequations}
	\begin{align}
		J_j(\vec{z}) & = \phi_{+j}(\vec{z}) - \phi_{-j}(\vec{z}) \\
		-\Delta_j G(\vec{z}) & = RT \ln \frac{\phi_{+j}(\vec{z})}{\phi_{-j}({\vec{z}})} \, ,
	\end{align}
	\label{eq:MAK}
\end{subequations}
\noindent where \(\phi_{+j}(\vec{z})\) (resp. \(\phi_{-j}(\vec{z})\)) is the precise rate at which the forward \(+j\) (resp. reverse \(-j\)) reaction occurs.
For example, for elementary reactions in ideal solutions, the reaction rates follow the mass--action kinetics:
\(\phi_{+j}(\vec{z}) = k_{+j} \prod_{i} z_{i}^{\nu_{i,j}}\) and \(\phi_{-j}(\vec{z}) = k_{-j} \prod_{i} z_{i}^{\bar\nu_{i,j}}\), where \(k_{+j}\) and \(k_{-j}\) are the rate constants.
When \Eqs{MAK} are satisfied, it is easy to see that \(J_j\) and \(- \Delta_j G\) are aligned for any \(\vec{z}\), and hence \(- J_j \, \Delta_j G \ge 0\).
Since any chemical process can be described as a network of elementary chemical reactions, the aforementioned fact constitutes a proof that \(\sigma \ge 0\).

Any thermodynamically consistent description ensures that when the system is closed, \(\vec{I}=0\), the steady state corresponds to thermodynamic equilibrium where all reaction currents vanish, \(\vec{J}(\vec{z})=\vec{0}\), as well as the dissipation, $\sigma=0$.
Indeed, the local detailed conditions (\ref{eq:MAK}) do so.
In the special case of mass--action kinetics, they also imply Wegscheider's conditions~\cite{Horn.Jackson1972,Schuster.Schuster1989}).

\subsection{Network topology and dissipation}
\label{sec:cycle-structure}

We now make use of the topology of the CRN encoded in its stoichiometric matrix.
To that end, we split the chemical concentrations into two groups: \(\vec{z} = (\vec{x}, \vec{y})\).
The species \(\ce{X}\) are entirely \emph{internal} to the reaction network and cannot be exchanged with the outside world, \textit{i.e.} \(\vec{I}^{\ce{X}}=0\).
Consequently, their dynamics are entirely due to reactions:
\begin{align}
  \dot{\vec{x}} &= \stoich^{\ce{X}} \, \vec{J}(\vec{x}, \vec{y})\,.
\end{align}
In contrast, the \emph{chemostatted} species \(\ce{Y}\) are controlled by the environment:
\begin{align}
  \dot{\vec{y}} &= \stoich^{\ce{Y}} \, \vec{J}(\vec{x}, \vec{y}) + \vec{I}^{\ce{Y}}\,. 
  \label{Beq_for_Y}
\end{align}
Such control can result from external reservoirs fixing their chemical potentials (or equivalently their concentrations in an ideal solution) \cite{avanzini21} or from mechanisms fixing their in- or out-fluxes \(\vec{I}^{\ce{Y}}\) \cite{avanzini22}.
Note that the splitting of the species results in a corresponding splitting of the stoichiometric matrix \( \stoich=(\stoich^{\ce{X}}, \stoich^{\ce{Y}})\T \).

At steady state, these two chemostatting procedures coincide and the reaction currents satisfy
\begin{align}
	0 &= \stoich^{\ce{X}}\,\vec{J}^{\ast}\,, \\
	0 &= \stoich^{\ce{Y}}\,\vec{J}^{\ast} + \vec{I}^{\ce{Y}}\,.
\end{align}
The steady-state currents thus need to be in the null-space of the reduced stoichiometric matrix \(\stoich^{\ce{X}}\): \(\vec{J}^{\ast} \in  \ker \stoich^{\ce{X}}\) at steady state.
Importantly, this space contains the null-space of \(\stoich\), \(\ker \stoich \subseteq \ker \stoich^{\ce{X}}\), whose elements represent combinations of reaction currents that when taken together neither produce nor consume any molecule:
they return the system back to its initial state.
We call these \emph{internal cycles} and let \(\set{\vec{c}^\ell}\) denote a maximal independent set of these---\textit{i.e.} a basis of \(\ker \stoich\).
The null-space \(\ker\stoich^{\ce{X}}\) is spanned by all combinations of reactions that leave the concentrations of internal species unvaried. 
This encompasses all internal cycles, but also accounts for additional cycles:
The steady state must exchange chemostatted species in order to faithfully represent a nonequilibrium state.
We call these \emph{additional} cycles \emph{emergent}, and denote a maximal independent set of them by \(\set{\vec{C}^\varepsilon}\).
Their stoichiometry is given by \(\stoich \vec{C} = (\vec{0}, \stoich^{\ce{Y}}\vec{C})\).
The internal species do not appear and only the chemostatted species remain. 
The positive terms are products and the negative terms are reactants.
Every emergent cycle thus defines an \emph{effective reaction} amongst the chemostatted species, which is fully balanced since each reaction in the network it was derived from was fully balanced.
Altogether they characterize the thermodynamic transformations that the CRN performs on the environment.

The thermodynamic importance of internal and emergent cycles becomes apparent when decomposing the steady-state currents \(\vec{J}^{\ast} = \sum_\ell J_\ell \vec{c}^\ell + \sum_{\varepsilon} J_\varepsilon \vec{C}^\varepsilon \) according to internal and emergent cycles.
Using this decomposition, the dissipation simplifies to
\begin{align}
	T\sigma = - \sum_j J^*_j \Delta_j G = - \sum_{\varepsilon} J_{\varepsilon} \Delta_{\varepsilon} G \, .
	\label{eq:EPssEmergent}
\end{align}
In the last equality, we exploited the fact that the internal cycles forces vanish
\begin{align}
  - \sum_j c_j \,\Delta_j G
  = - \vec{c} \stoich\T \vec{\mu}
  = - \left(\stoich \vec{c} \right)\T \vec{\mu} = 0 \, .
\end{align}
Therefore, while the internal cycle currents \(J_\ell\) do not vanish in general, they do not produce dissipation.
Instead, the emergent cycle forces 
\begin{align}
  -\Delta_\varepsilon G = -\sum_j \vec{C}^{\varepsilon}_j \Delta_j G =- \sum_j \vec{C}_j^\varepsilon \stoich\T_j \vec{\mu} = (\stoich^{\ce{Y}}\vec{C}^{\varepsilon})\T \vec{\mu}^{\ce{Y}} \, ,
\end{align} 
typically do not vanish.
They only involve the chemical potentials of the chemostatted species since the contributions of internal species, by construction, cancel in the summation. 

The dissipation in a nonequilibrium steady state is thus entirely captured by the currents and forces along the emergent cycles \cite{Polettini.Esposito2014}.
From a thermodynamic standpoint they thus constitute the fundamental set of currents and forces in the entire open CRN.

We emphasize that our framework holds far-from-equilibrium in regimes where the fluxes are nonlinear functions of the forces.
Linear irreversible thermodynamics is recovered close to equilibrium, where the fluxes become linear in the forces \(J_{\varepsilon} \approx \sum_{\varepsilon'} \mathcal{L}_{\varepsilon \varepsilon'} \Delta_{\varepsilon'} G\), where $\mathcal{L}$ is the Onsager matrix of the open CRN.

\section{Free-energy transduction in simple models}
\label{sec:coupling}

Before proceeding, we define a \emph{chemical process} as a generic stoichiometric relation between reactants and products, like \Eq{crn:stoichiometry}.
Such a stoichiometric relation must be balanced with respect to electrons and atomic nuclei---and thus charge and mass---but the path leading from reactants to products can be arbitrary.
In this way, a chemical process is defined independently from the kinetics that characterises its rate of occurrence.
The stoichiometry of the processes is however sufficient to determine its thermodynamic force via \Eq{crn:force}, since chemical potentials are state functions.

An example of a chemical process is the oxidation of glucose:
\begin{align}
	\ce{Glu + 6 O2} \rightleftharpoons \ce{6 CO2 + 6 H2O} \, .
\end{align}
This process can be the result of combustion or of cellular respiration.
While the kinetics is different in these two cases, we still consider it to be the same chemical process with the same thermodynamic force.
An emergent cycle \(\vec{C}^\varepsilon\) also defines a chemical process.

We now have all the ingredients at hand to characterize the free-energy transduction of chemical processes.
In this section we start by discussing simple models that may describe enzymatic mechanisms or membrane transport processes.
In the next section, we will scale up our analysis to metabolic processes.

\subsection{Independent processes}
\label{sec:independent}

\begin{figure}[h]
	\centering
	\includegraphics[scale=0.8]{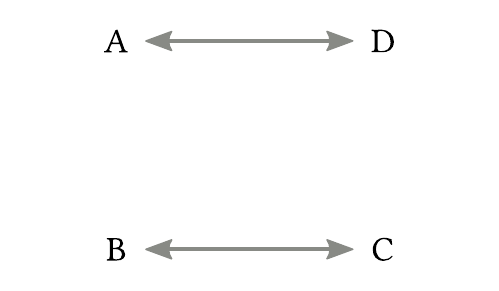}
	\includegraphics[scale=0.8]{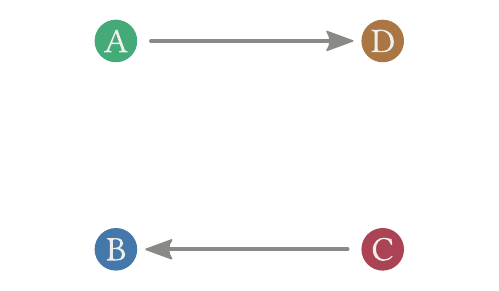}
	\caption{
		(Left) Closed CRN made of two \emph{independent} reactions.
		(Right) The two emergent cycles that arise when the CRN is opened by chemostatting the species \ce{A, B, C,} and \ce{D}.
	}
	\label{fig:independent}
\end{figure}

Let us consider the CRN 
\begin{align}
  \ce{A} \rightleftharpoons \ce{D} \ \ \ \ \ \ \ \ \ce{B} \rightleftharpoons \ce{C}
  \label{eq:tc--network_Indep}
\end{align}
depicted in Fig.~\ref{fig:independent} (left).
Each reaction could consist of many intermediate steps, as long as these are \emph{independent} from the other reaction.
Furthermore, since each reaction must preserve mass, they can be thought of as different conformations of a molecule.
The CRN is opened by chemostatting the species A, B, C, and D as depicted in Fig.~\ref{fig:independent} (right).
The two emergent cycles in this system are identical to the two reactions (\ref{eq:tc--network_Indep}).

We now consider the two independent chemical processes:
\(\pf\colon \ce{A} \rightleftharpoons \ce{D}\) and
\(\ps\colon \ce{B}\rightleftharpoons \ce{C}\).
We note that in this trivial example, reactions, emergent cycles and processes all coincide.
These two processes happen with net currents $J\supsc{ind}_{\pf}$ and $J\supsc{ind}_{\ps}$, and their direction is determined by the external forces,
\begin{align}
	-\Delta_{\pf} G = \mu_{\ce{A}} - \mu_{\ce{D}} \, ,  \quad
	-\Delta_{\ps} G = \mu_{\ce{B}} - \mu_{\ce{C}} \, .
\end{align}
Both processes reach thermodynamic equilibrium if and only if \(\mu_{\ce{A}} = \mu_{\ce{D}}\) and \(\mu_{\ce{B}} = \mu_{\ce{C}}\).

The total entropy-production rate is the sum of the rates associated with each of the two emergent cycles, which coincide with the two independent processes:
\begin{align}
	T \sigma\supsc{ind} = -J\supsc{ind}_{\pf} \Delta_{\pf} G - J\supsc{ind}_{\ps} \Delta_{\ps}G \, .
\end{align}
The assumption of independence between the two reactions allows us to state that each of the processes needs to satisfy the second law independently.
Hence,
\begin{align}
	T\sigma\supsc{ind}_{\pf} \coloneqq -J\supsc{ind}_{\pf} \Delta_{\pf} G \geq 0 \, , \quad
	T\sigma\supsc{ind}_{\ps} \coloneqq -J\supsc{ind}_{\ps} \Delta_{\ps} G \geq 0 \, .
	\label{eq:independent-directions}
\end{align}
This makes the two independent processes purely dissipative, and implies that each reaction current is following the direction of its thermodynamic force.
Free-energy transduction cannot happen in this case.

We will now consider different ways in which these two processes might be coupled resulting in transduction.

\subsection{Tightly coupled processes}

\begin{figure}[h]
	\includegraphics[scale=0.8]{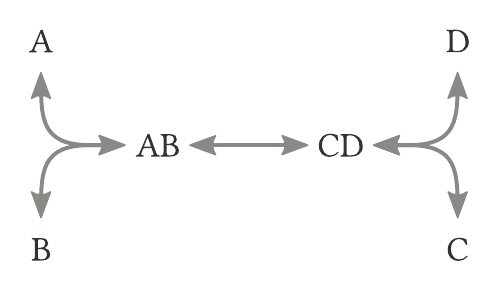}
	\includegraphics[scale=0.8]{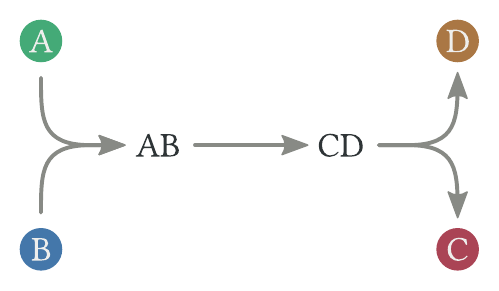}
	\caption{
		(Left) Closed CRN made of three reactions.
		(Right) The single emergent cycle that arises when \ce{A}, \ce{B}, \ce{C}, and \ce{D} are chemostatted.
		This CRN implements the same two processes as in Fig.~\ref{fig:independent}, but with intermediate steps that couple the two processes \emph{tightly}, \textit{i.e.} via a fixed stoichiometry.
	}
	\label{fig:tight}
\end{figure}

Let us assume that the same four species \ce{A}, \ce{B}, \ce{C}, and \ce{D} we introduced in \S\ref{sec:independent} are the only chemostats in the following CRN with arbitrary kinetics:
\begin{align}
  \ce{A + B} \rightleftharpoons \ce{AB} \rightleftharpoons \ce{CD} \rightleftharpoons \ce{C + D} \, .
  \label{eq:tc--network}
\end{align}

We consider the same two processes as before:
\(\pf\colon \ce{A} \rightleftharpoons \ce{D}\) and \(\ps\colon \ce{B} \rightleftharpoons {C}\).
This time the CRN has only one emergent cycle with stoichiometry
\begin{align}
	\ce{A + B} \rightleftharpoons \ce{C + D} \, ,
	\label{eq:tc--net-stoich}
\end{align}
as illustrated in Fig.~\ref{fig:tight} to highlight the difference with respect to the previous case.

The two individual processes \(\pf\colon \ce{A} \rightleftharpoons \ce{D}\) and \(\ps\colon \ce{B} \rightleftharpoons {C}\) come now bound with a fixed stoichiometric ratio of unity: a molecule of type \ce{D} can only be created when also a molecule of type \ce{C} is created.
Therefore, we say that the two processes are \emph{tightly coupled}.
The force of the emergent cycle is the sum of the individual forces:
\begin{align*}
	-\Delta\subsc{tight} G
	= \mu_{\ce{A}} + \mu_{\ce{B}} - \mu_{\ce{C}} - \mu_{\ce{D}}
	= -\Delta_{\pf}G - \Delta_{\ps} G \, .
\end{align*}
Let \(J\supsc{tight}\) be the current of the emergent cycle in the steady state, then the entropy-production rate is
\begin{align}
  T\sigma = - J\supsc{tight} \Delta\subsc{tight}G = - J\supsc{tight} \Delta_{\pf}G - J\supsc{tight}\Delta_{\ps}G\,.
  \label{eq:tc--epr-decomposition}
\end{align}
This expression is, in total, always non-negative: \(T\sigma \geq 0\).
Hence, we can deduce that either
\begin{enumerate}
	\item \(-\Delta_{\pf}G = \Delta_{\ps} G\) and the system is at equilibrium, \(J\supsc{tight}=0\)\,.
	\item \(-\Delta_{\pf} G, -\Delta_{\ps}G < 0\) or \(-\Delta_{\pf} G, -\Delta_{\ps}G > 0\) which makes the system purely dissipative.
	\item \(-\Delta_{\pf} G\) and \(-\Delta_{\ps}G\) have different sign.
		Then the system can be understood to transduce free energy.
		In the following we assume, without restriction, that \(-\Delta_{\pf}G\,,J\supsc{tight} >0\) while \(-\Delta_{\pf}G > \Delta_{\ps}G > 0\).
\end{enumerate}
In the latter situation, we can say that the tight coupling mechanism uses the free energy released by the process \(\pf\) as input to drive the (output) process \(\ps\) uphill in free energy:
\begin{align}
	T\sigma\supsc{tight}_{\pf} \coloneqq -J\supsc{tight}\Delta_{\pf}G > 0 \, , \quad
	T\sigma\supsc{tight}_{\ps} \coloneqq -J\supsc{tight}\Delta_{\ps} G < 0 \;.
\end{align}

This example represents the simplest form of a chemical machine.
The machine itself is given by the internal species \ce{AB} and \ce{CD} and operates on an environment given by the chemostats \ce{A}, \ce{B}, \ce{C}, and \ce{D}.
The efficiency of free-energy transduction for this machine is the ratio of the output over the input free energy,
\begin{align}
  \eta\supsc{tight} \coloneqq - \frac{T\sigma\supsc{tight}_{\ps}}{T\sigma\supsc{tight}_{\pf}} = \frac{\Delta_{\ps}G}{-\Delta_{\pf}G} \leq 1 \,.
  \label{eq:tc--efficiency}
\end{align}
This efficiency is bounded from above by unity, since \(T\sigma\supsc{tight} = T\sigma\supsc{tight}_{\pf} + T\sigma\supsc{tight}_{\ps} \geq 0\).
Maximal efficiency is achieved only at equilibrium, where both \(J\supsc{tight}=0\) and \(-\Delta\subsc{tight} G = 0\) wich implies \(T\sigma\supsc{tight}_{\pf} = T\sigma\supsc{tight}_{\ps} = 0\) while \(-\Delta_{\pf}G = \Delta_{\ps}G\) may be arbitrarily large.

Also note that \(\eta\supsc{tight}\) is entirely expressed in terms of the thermodynamic forces.
The kinetics of the coupling process only affect the current, which cancels out.
This is the distinctive feature of tightly coupled processes.

\subsection{Loosely coupled processes}

\begin{figure}[h]
	\includegraphics[scale=1.0]{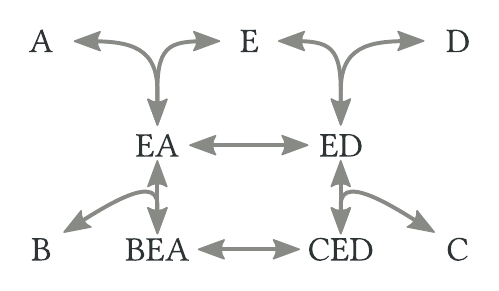}
	\caption{
		Closed CRN describing two loosely coupled processes mediated by an enzymatic mechanism.
		The open CRN obtained by chemostatting \ce{A, B, C} and \ce{D} (Fig.~\ref{fig:loosely-cycles}) is characterized by two linearly independent emergent cycles.
		One choice of basis for the emergent cycles is depicted in Fig.~\ref{fig:loosely-cycles}.
	}
	\label{fig:loosely}
\end{figure}

A different form of coupling can be achieved by considering some form of catalyst, \(\ce{E}\), that couples the conversion of \(\ce{A}\) and \(\ce{B}\) into \(\ce{D}\) and \(\ce{C}\) using the CRN:
\begin{align}
  \ce{A + E} \rightleftharpoons \ce{EA} &\rightleftharpoons \ce{ED} \rightleftharpoons \ce{E + D} \\
  \ce{B + EA} \rightleftharpoons \ce{BEA} &\rightleftharpoons \ce{CED}
  \rightleftharpoons \ce{C + ED}
\end{align}
also depicted in Fig.~\ref{fig:loosely}.

With the species \ce{A}, \ce{B}, \ce{C}, and \ce{D} chemostatted, we have two emergent cycles and we have to choose a basis.
The three possible choices are:
\begin{itemize}
  \item \( \ce{A} \rightleftharpoons \ce{D}\) and
    \( \ce{B}\rightleftharpoons \ce{C}\)
  \item \( \ce{A + B} \rightleftharpoons \ce{C + D}\) and
    \( \ce{B}\rightleftharpoons \ce{C}\)
  \item \( \ce{A + B} \rightleftharpoons \ce{C + D}\) and
    \( \ce{A}\rightleftharpoons \ce{D}\)
\end{itemize}

\begin{figure}[h]
  \includegraphics[scale=0.8]{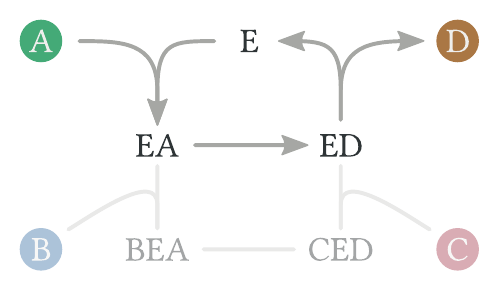}
  \includegraphics[scale=0.8]{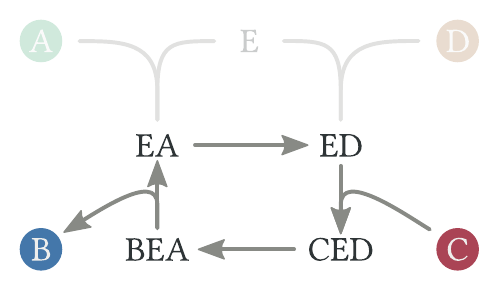}
  \caption{The two coupled emergent cycles \( \ce{A} \rightleftharpoons \ce{D}\) and \( \ce{B}\rightleftharpoons \ce{C}\) of the catalytic reaction network shown in Fig.~\ref{fig:loosely}.}
  \label{fig:loosely-cycles}
\end{figure}

Since we are still interested in the chemical processes \(\pf\colon \ce{A} \rightleftharpoons \ce{D}\) and
\(\ps\colon \ce{B}\rightleftharpoons \ce{C}\)---without loss of generality---we choose the first choice of emergent cycles, which are depicted in Fig.~\ref{fig:loosely-cycles}.
In this way each emergent cycle carries a process. 
These cycles have the same forces as the independent ones in section~\ref{sec:independent}.
The fact that there is a kinetic coupling does not affect the thermodynamic equilibrium, nor the thermodynamic forces.
However, two currents are needed now to characterize dissipation:
\begin{align}
	T\sigma\supsc{cat} &= -J\supsc{cat}_{\pf} \Delta_{\pf} G - J\supsc{cat}_{\ps} \Delta_{\ps}G \\
	&= T\sigma\supsc{cat}_{\pf} + T\sigma\supsc{cat}_{\ps} \, .
\end{align}
Note that the two currents are \emph{not} the currents coming from the kinetics of the independent processes used in \S\ref{sec:independent}, but from the catalytic coupling and thus depend on the total amount of available catalyst.

Again, since the overall dissipation is non-negative, we have three cases:
(1) thermodynamic equilibrium, (2) pure dissipation, or (3) transduction.
In the latter case we assume, without restriction, \(T\sigma\supsc{cat}_{\pf}> 0\) while \(T\sigma\supsc{cat}_{\ps} < 0\).
Then the efficiency of free-energy transduction is
\begin{align}
	\eta\supsc{cat} = -\frac{T\sigma\supsc{cat}_{\ps}}{T\sigma\supsc{cat}_{\pf}}
	= \frac{J\supsc{cat}_{\ps}\Delta_{\ps}G}{-J\supsc{cat}_{\pf}\Delta_{\pf}G}
	= \eta\supsc{tight} \frac{J\supsc{cat}_{\ps}}{J\supsc{cat}_{\pf}} \leq 1 \, .
	\label{eq:effLC}
\end{align}
This efficiency is different from the tightly coupled case: It depends on the kinetics via the currents.
In general, there is no way to tell which of the two efficiencies is larger.

In this example, the internal species all contain the catalyst \ce{E} in different states of binding to other molecules. The catalyst can thus be seen as the machine acting on an environment given by the chemostatted species.

\subsection{Hybrid cases}
\label{sec:hybrid}

\begin{figure}[h]
	\includegraphics[scale=1.0]{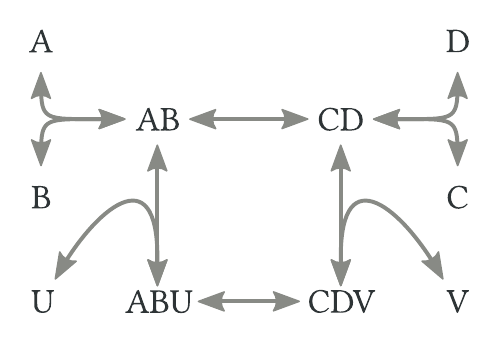}
	\caption{
		A closed CRN representing a hybrid between the two CRNs given in Figs.~\ref{fig:tight} and \ref{fig:loosely}.
		The intermediate steps of one conversion process acts as a catalyst for another.
		Chemostatting the six base species results in the two emergent cycles shown in Fig.~\ref{fig:hybrid-forward}.
	}
	\label{fig:hybrid-bidirectional}
\end{figure}

\begin{figure}[h]
  \includegraphics[scale=0.8]{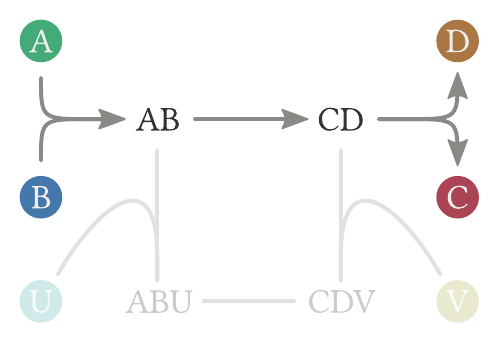}
  \includegraphics[scale=0.8]{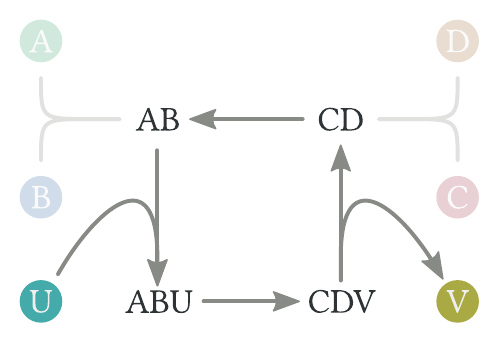}
  \caption{The two emergent cycles of the chemostatted network introduced in Fig.~\ref{fig:hybrid-bidirectional}.
  (left) Emergent cycle $\ef$ which tightly couples the processes \(\pf\) and \(\ps\).
  (right) Emergent cycle $\es$ which transforms \ce{U} into \ce{V}.}
  \label{fig:hybrid-forward}
\end{figure}

We now consider the CRN depicted in Fig.~\ref{fig:hybrid-bidirectional} that is a hybrid between the tightly and loosely coupled CRNs introduced earlier
\begin{align}
  \ce{A + B} \rightleftharpoons \ce{AB} &\rightleftharpoons \ce{CD} \rightleftharpoons \ce{C + D} \\
  \ce{AB + U} \rightleftharpoons \ce{ABU} &\rightleftharpoons \ce{CDV} \rightleftharpoons \ce{CD + V}
\end{align}
With the six chemostats \(\ce{A}\), \(\ce{B}\), \(\ce{C}\), \(\ce{D}\), \(\ce{U}\), \(\ce{V}\), we have two emergent cycles
\begin{itemize}
  \item \(\ef \colon  \ce{A + B} \rightleftharpoons \ce{C + D}\)
  \item \(\es \colon \ce{U} \rightleftharpoons \ce{V}\)
\end{itemize}
as depicted in Fig.~\ref{fig:hybrid-forward}, and the dissipation reads
\begin{align}
	T\sigma\supsc{hyb} = T\sigma\supsc{hyb}_{\ef} + T\sigma\supsc{hyb}_{\es} \, .
\end{align}

As before, we are interested in the chemical processes:
\(\pf\colon \ce{A} \rightleftharpoons \ce{D}\) and
\(\ps\colon \ce{B}\rightleftharpoons \ce{C}\) which are both tightly coupled by the emergent cycle $\ef$.
But contrary to the tight coupling case we now have an additional dissipative process transforming \ce{U} into \ce{V} on the emergent cycle $\es$. 

The dissipation can thus be expressed in terms of \emph{three} contributions:
\begin{align}
  T\sigma\supsc{hyb} = T\sigma\supsc{hyb}_{\pf} + T\sigma\supsc{hyb}_{\ps} + T\sigma\supsc{hyb}_{\es} \, .
\end{align}
Transduction arises when \(T\sigma\supsc{hyb}_{\ps} < 0\), but we need to distinguish two cases.
If the current of the additional process follows its force, \(T\sigma\supsc{hyb}_{\es} > 0\), it should be treated as an input process, and the efficiency is
\begin{align}
	\eta\supsc{hyb} &= - \frac{T\sigma\supsc{hyb}_{\ps}}{T\sigma\supsc{hyb}_{\pf} + T\sigma\supsc{hyb}_{\es}}
	= \frac{J\supsc{hyb}_{\ef}\Delta_{\ps}G}{-J\supsc{hyb}_{\ef}\Delta_{\pf}G -J\supsc{hyb}_{\es}\Delta_{\es}G  } \\
	&= \eta\supsc{tight} \left(1+\frac{J\supsc{hyb}_{\es}\Delta_{\es}G}{J\supsc{hyb}_{\ef}\Delta_{\pf}G}\right)^{-1} = \eta\supsc{tight} \left( 1 + \frac{\sigma\supsc{hyb}_{\es}}{\sigma\supsc{hyb}_{\pf}} \right)^{-1} \, ,
	\label{eq:effHyb}
\end{align}
where we used the fact that since the coupling between \(\pf\) and \(\ps\) is tight, their currents will be equal in strength.
Since $\sigma_{\es}>0$, the efficiency of transduction in this hybrid case is smaller than if we only had \(\pf\) tightly coupled to \(\ps\).
If instead \(T\sigma_{\ps} < 0\) and \(T\sigma_{\es} < 0\), then \(\es\) should be treated as an additional output process, and the efficiency should be defined as 
\begin{align}
	\eta\supsc{hyb'} = - \frac{T\sigma_{\ps} + T\sigma_{\es}}{T\sigma_{\pf}} =  \eta\supsc{tight} - \frac{\sigma_{\es}}{\sigma_{\pf}}\, .
\end{align}
In this case, since multiple output processes are obtained from the same input, the efficiency increases with respect to the tightly coupled case.

\subsection{Counter-Example to Transduction: Mass transfer}
\label{sec:massTransfer}

\begin{figure}[h]
  \includegraphics[scale=1.0]{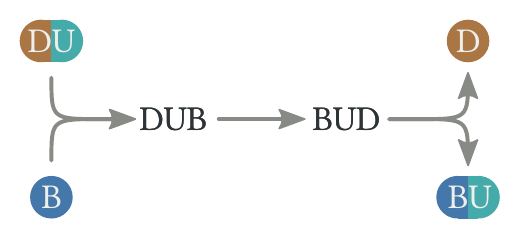}
  \caption{Mass transfer is a counter example to transduction.}
  \label{fig:mass-transfer}
\end{figure}

In order to define transduction, it is crucial to to make sure that the two processes---inputs and outputs---are fully balanced in terms of electrons and atomic nuclei.
Here we highlight this fact.

Let us re-consider the example of \emph{tight coupling} as given in \Eq{tc--network} or Fig.~\ref{fig:tight}.
Only looking at this reaction network or its emergent cycle stoichiometry \Eq{tc--net-stoich} cannot guarantee that the species \(\ce{A}\) and \(\ce{D}\) have the same charge and atomic composition, nor that there is an actual process that could transform \(\ce{A}\) into \(\ce{D}\).
Let us assume for instance that \(\ce{A} = \ce{DU}\) and \(\ce{C}=\ce{BU}\), meaning that the net stoichiometry of the emergent cycle would be
\begin{align}
  \ce{DU} + \ce{B} \rightleftharpoons \ce{BU} + \ce{D}\,,
  \label{eq:counter-example--process}
\end{align}
as depicted in Fig.~\ref{fig:mass-transfer}.
This represents the net transfer of the moiety \(\ce{U}\) from \(\ce{D}\) to \(\ce{B}\).
In this case, there is no way to decompose this process into two independent chemical processes and to naturally decompose the entropy-production rate as we did in \Eq{tc--epr-decomposition}.
Mass transfer does not correspond to free-energy transduction.

\subsection{Nonlinear Case Study}
\label{sec:nonlinear}

\begin{figure}[h]
  \includegraphics[scale=1.]{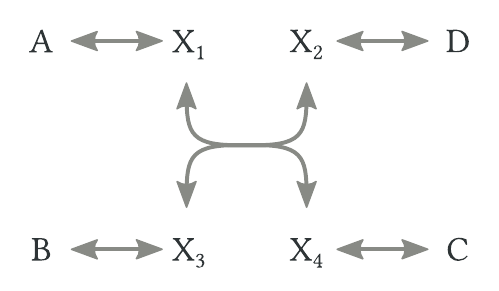}
  \caption{
	  Nonlinear closed CRN given by the reactions (\ref{Lreact}) and (\ref{NLreact}).
  }
  \label{fig:nonlinear-example}
\end{figure}

The models discussed so far are pseudo-unimolecular: when considering the chemostatted species constant in time, all reactions are unimolecular and dynamically linear in the internal species.
But the theory in section~\ref{sec:chem-networks}, as well as our coming main result in section~\ref{sec:transduction}, do not rely on this fact. 

\begin{figure}[h]
  \includegraphics[scale=1.]{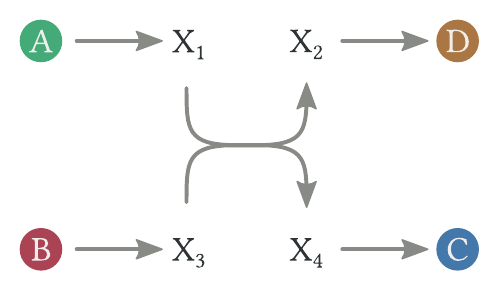}
  \caption{Emergent cycle \(\ce{A}+\ce{B} \rightleftharpoons \ce{C} + \ce{D}\) obtained when chemostatting (\ce{A}, \ce{B}, \ce{C}, \ce{D}) in the CRN of Fig. \ref{fig:nonlinear-example}.}
  \label{fig:nonlinear-cycle}
\end{figure}

To illustrate this point, we consider the non-linear CRN
\begin{align}
   \ce{A} \rightleftharpoons \ce{X}_1 \ \; \ \ce{B} \rightleftharpoons \ce{X3} \ \; & \ 
   \ce{C} \rightleftharpoons \ce{X}_1  \ \; \ \ce{D} \rightleftharpoons \ce{X}_2 \label{Lreact} \\
   \ce{X}_1 + \ce{X}_2 &\rightleftharpoons \ce{X}_3 + \ce{X}_4 \label{NLreact}
\end{align}
depicted in Fig.~\ref{fig:nonlinear-example}.
Its central reaction \eqref{NLreact} takes two internal species into two different internal species, making it dynamically nonlinear even when the chemostats \ce{A}, \ce{B}, \ce{C}, and \ce{D} have fixed concentrations.
Indeed, the chemostatting produces the single emergent cycle \(\ce{A}+\ce{B} \rightleftharpoons \ce{C} + \ce{D}\) depicted in Fig.~\ref{fig:nonlinear-cycle}.

This example provides an alternative way to tightly couple the processes \(\pf\) and \(\ps\).
From the perspective of the reservoirs, it is essentially identical to the system in Fig.~\ref{fig:tight}.
However, the nonlinear nature of its internal dynamics will have an impact on the global kinetics of the system:
The total current through this system has a different dependence on the chemostats' concentrations than the linear system discussed earlier.
Thus, the global kinetics can be a way to differentiate the two different implementations of tight coupling.

\section{Free-energy transduction in central metabolism}
\label{sec:metabolism}

    We now apply the framework introduced in the previous sections to a biochemically relevant case: central energy metabolism.
The function of central energy metabolism is the transfer of energy from energy-rich molecules---most importantly glucose---into \(\ce{ATP}\) \cite{nelson08,voet16}.

We first consider cellular respiration and its three components: glycolysis, the tricarboxylic acid (TCA) cycle, and the electron transport chain with ATP synthesis.
We show that the corresponding metabolic pathways are recovered as emergent cycles, and hence they can be interpreted as chemical processes.
From the perspective of free-energy transduction, we find that all these energy-conversion pathways are tightly coupled.
This allows us to estimate the efficiency of these chemical machines from the \(\Delta G\)s alone.

We then conclude our analysis by considering two extensions of glycolysis: fermentation and gluconeogenesis.
While the former can be interpreted as a tightly-coupled process, the latter cannot, being composed of several separate processes.

We refer to \App{CentralEnergyMetabolism} for a detailed discussion of all individual reaction steps.

\subsection{Cellular Respiration}

With its high yield of \ce{ATP}, cellular respiration plays a central role in energy metabolism, and is shared among most cellular life forms with essentially the same structure.
The three components of respiration are glycolysis in which glucose is converted into pyruvate while synthesizing \ce{ATP}s, TCA cycle in which pyruvate is used to reduce the coenzymes \ce{CoQ} and \ce{NAD+}, and electron transport chain in which the reducing power of \ce{NADH} is used to synthesize \ce{ATP}. We will be discussing these pathways individually in the following sections.

\begin{figure*}[p]
    \centering
    \includegraphics[width=0.85\textwidth]{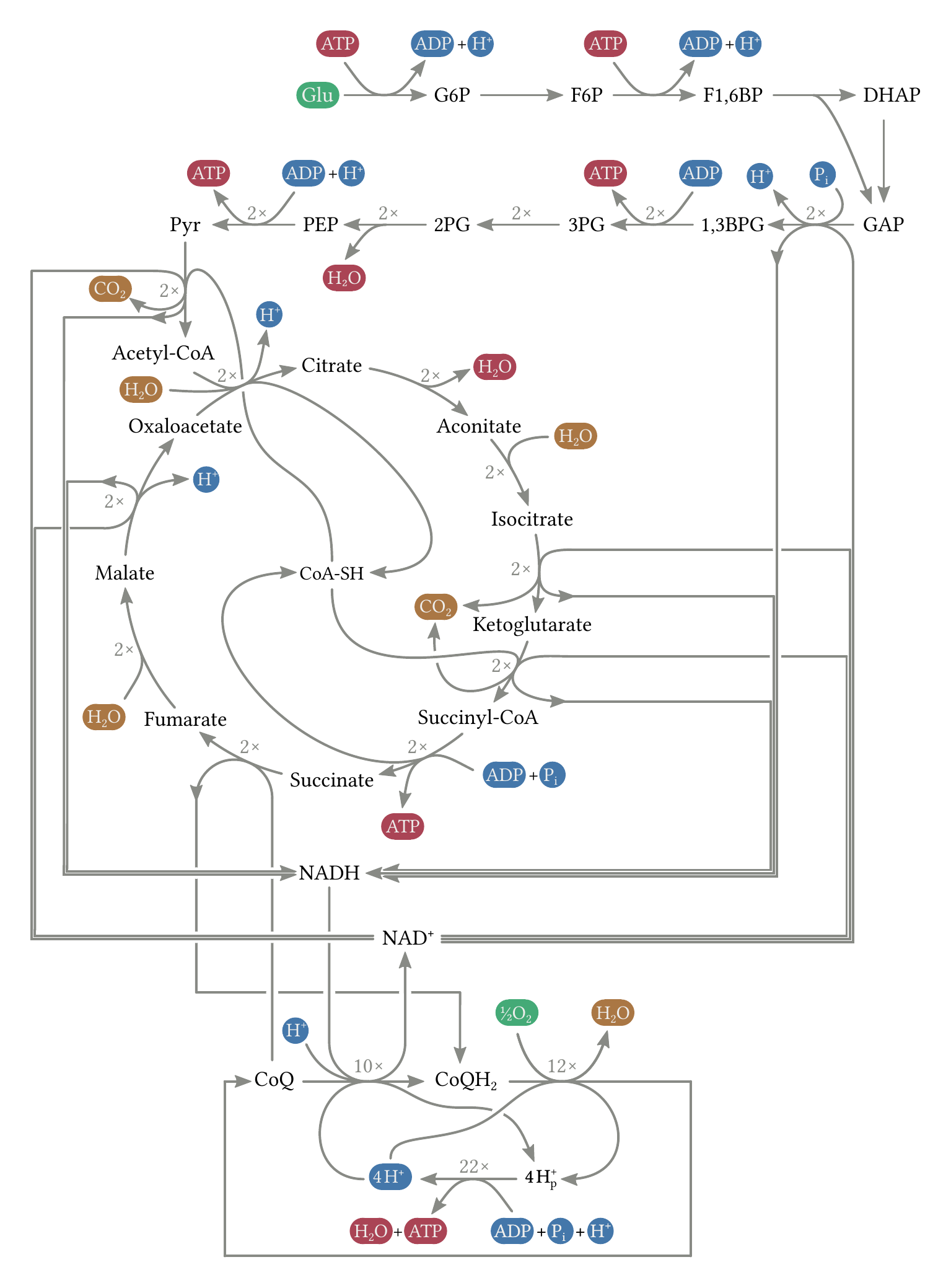}
    \caption{
        The single emergent cycle of \textbf{cellular respiration} is formed by a combination of glycolysis, the TCA cycle, as well as the electron transport chain.
        Each reaction has indicated multiplicities, so as to generate the net balance given in (\ref{eq:respiration:net-balance}).
        The coloring highlights our interpretation of two coupled processes in the net reaction:
        red/blue is the synthesis of \ce{ATP} and green/brown is the breakdown of glucose (\ce{Glu}) into two molecules of pyruvate (\ce{Pyr}).
    }
    \label{fig:energy-metabolism}
\end{figure*}

Importantly, when all the reactions of these three pathways (see (\ref{eq:glycolysis--reactions}a--j), \eqref{eq:TCA:reactions} and  \eqref{eq:electron-transport-chain--pathway}) are taken together as part of one single network and when \ce{H2O}, \ce{H+}, \ce{ATP}, \ce{ADP}, \ce{P_{\textrm{i}}}, \ce{O2}, \ce{CO2}, \ce{Glu} are regarded as chemostatted, then one observes the single emergent cycle
\begin{gather}
	\ce{Glu + 6 O2 + 26 ADP + 26 P_{\mathrm{i}} + 26 H+} \nonumber \\
	\updownharpoons \label{eq:respiration:net-balance} \\
	\ce{6 CO2 + 32 H2O + 26 ATP}   \nonumber
\end{gather}
depicted in Fig. \ref{fig:energy-metabolism}. 
The choice of the chemostatted species is suggested by viewing cellular respiration as a cellular process operating on the environment.
\ce{Glu}, \ce{O2}, \ce{CO2}, \ce{H2O}, and \ce{H+} are chemicals that cells exchange with the environment, whereas the synthesis of \ce{ATP} from \ce{ADP} and \ce{P_{\textrm{i}}} is the desired cellular chemical output.

Following this reasoning and using the information about the balance of charges and atomic nuclei, we recognize that the net stoichiometric balance of cellular respiration is the result of two chemical processes tightly coupled together:
the burning of glucose into carbon dioxide and the synthesis of ATP,
\begin{align}
	& \ce{Glu + 6 O2} \rightleftharpoons \ce{6 CO2 + 6 H2O} \label{eq:GluOxidat} \\
	& \ce{ADP + P_{\mathrm{i}} + H+} \rightleftharpoons \ce{ATP + H2O} \label{eq:ATPsynth} \, ,
\end{align}
The stoichiometric coupling of 26 \ce{ATP} per \ce{Glu} in central energy metabolism depends on the species \cite{voet16}.
Here we consider the bacterium \emph{E. coli}.
For eukaryotes, the organization of central energy metabolism is slightly different: for them the TCA cycle and the electron transport chain are performed in the mitochondria.
The transport into and out of mitochondria is partially active and thus itself consumes some \(\ce{ATP}\).
Bacteria do not have specialized organelles for this function.
This difference, together with sightly different enzymes for the electron transport chain, results in a different global stoichiometry.

Under standard physiological conditions (see \App{CentralEnergyMetabolism}), the full oxidation of glucose \eq{GluOxidat} releases about \SI{2910}{\kilo\joule\per\mol} ($1180 \, RT$), while the free energy of the \(\ce{ATP}\) hydrolysis reaction is \(\SI{46}{\kilo\joule\per\mole}\) ($18.4 \, RT$).
Thus, the theoretical (physical) limit to the conversion rate is roughly 64 \(\ce{ATP}\) per 1 glucose.
Considering that cellular respiration produces 26 \(\ce{ATP}\) per glucose, this makes for an efficiency of about 41\%:
\begin{equation}
	\eta_{\mathrm{CR}} = \frac{26 \, \Delta_{\mathrm{ATPsynth}}G}{- \Delta_{\mathrm{GluOxidat}}G} = 0.41 \, ,
	\label{eq:CR:eff}
\end{equation}
where
\begin{align}
	- \Delta_{\mathrm{GluOxidat}}G & = \mu_{\ce{Glu}} + 6 \mu_{\ce{O2}} - 6 \mu_{\ce{CO2}} - 6 \mu_{\ce{H2O}} \label{eq:GluOxidat:DeltaG} \\
	& = \SI{2910}{\kilo\joule\per\mol} \notag \\
	- \Delta_{\mathrm{ATPsynth}}G & = \mu_{\ce{ADP}} + \mu_{\ce{Pi}} + \mu_{\ce{H+}} - \mu_{\ce{ATP}} - \mu_{\ce{H2O}} \label{eq:ATPsynth:DeltaG} \\
	& = - \SI{46}{\kilo\joule\per\mol} \, . \notag
\end{align}

We now turn to the individual analysis of the three pathways constituting cellular respiration.

\subsubsection{Glycolysis}

\begin{figure*}[tbh]
	\vspace{-0.5cm}
	\includegraphics[scale=0.9]{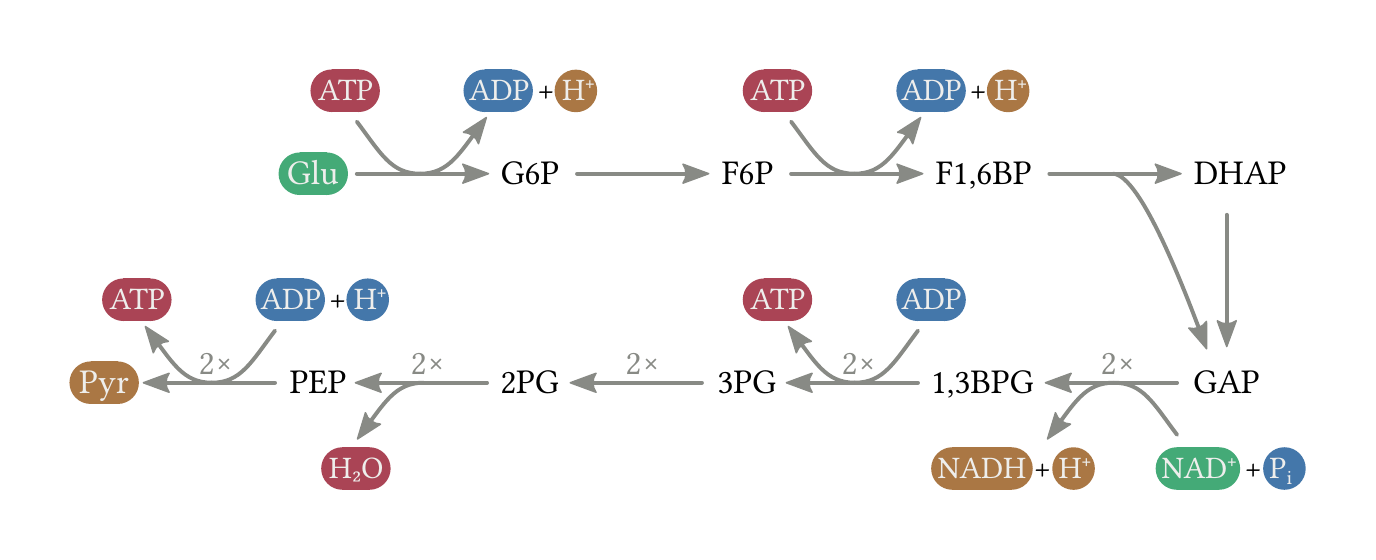}
	\vspace{-0.5cm}
	\caption{
	Single emergent cycle of \textbf{glycolysis} which couples the synthesis of \ce{ATP} (red/blue) and the breakdown of glucose into two molecules of pyruvate (green/brown).
	}
	\label{fig:glyco:open:forward}
\end{figure*}

Glycolysis is the first part of cellular respiration, and its individual steps are the enzymatic reactions depicted in Fig.~\ref{fig:glyco:open:forward}, see also Eqs.~(\ref{eq:glycolysis--reactions}a--j).
The chemostatting of \ce{Glu}, \ce{Pyr}, \ce{ATP}, \ce{ADP}, \ce{P_{\textrm{i}}}, \ce{NAD+}, \ce{NADH}, \ce{H+} and \ce{H2O}, produces a single emergent cycle, which re-traces the traditional understanding of the glycolysis pathway.
It begins with one molecule of glucose (\ce{Glu}) and ends with two molecules of pyruvate (\ce{Pyr}), producing a net gain of 2 \ce{ATP} per glucose.
In order to realize this process, the glucose is phosphorylated twice before it is broken into two 3-carbon bodies, \ce{DHAP} and \ce{GAP}.
These can then be easily converted into each other.
In \ce{GAP} form, the two pieces are phosphorylated again via the help of \ce{NAD+}, which is reduced in the process.
Finally, each of the carbon bodies is de-phosphorylated twice, resulting in the total production of 4 \ce{ATP}, leaving a net gain of 2 \ce{ATP} per glucose.

The net balance of glycolysis coincides with the effective reaction that the emergent cycle performs on the chemostats 
\begin{gather}
  \ce{Glu + 2 ADP + 2 P_{\mathrm{i}} + 2 NAD+}  \nonumber\\
  \updownharpoons \label{eq:glycolysis-net-stoichiometry}\\
  \ce{2 Pyr + 2 ATP + 2 NADH + 2 H+ + 2 H2O}\,. \nonumber
\end{gather}
We can easily identify the synthesis of \ce{ATP} from \ce{ADP} as one of the processes involved, \Eq{ATPsynth}.
The other process is instead the conversion of glucose into pyruvate:
\begin{align}
	\ce{Glu + 2 NAD+} &\rightleftharpoons \ce{2 Pyr + 2NADH + 4H+} \label{eq:Glu2Pyr}
\end{align}

Under standard physiological conditions (details in \App{CentralEnergyMetabolism}) the conversion of glucose into pyruvate, \Eq{Glu2Pyr}, releases roughly \(\SI{170}{\kilo\joule\per\mole}\) ($68 \, RT$).
Recalling that (i) the free energy required for ATP synthesis, \Eq{ATPsynth:DeltaG}, is \(\SI{46}{\kilo\joule\per\mole}\) ($18.4 \, RT$), and (ii) two ATPs are hydrolyzed during glycolysis, one can estimate the efficiency of this pathway,
\begin{equation}
	\eta_{\mathrm{glyco}} = \frac{2 \, \Delta_{\mathrm{ATPsynth}}G}{- \Delta_{\ce{Glu}\rightarrow\ce{Pyr}}G}
	= 0.54 \, ,
	\label{eq:glyco:eff}
\end{equation}
with
\begin{align}
	\hspace{-1em}
	- \Delta_{\ce{Glu}\rightarrow\ce{Pyr}}G & = \mu_{\ce{Glu}} + 2 \mu_{\ce{NAD+}} - 2 \mu_{\ce{Pyr}} - 2 \mu_{\ce{NADH}} - 4 \mu_{\ce{H+}} \label{eq:Glu2Pyr:DeltaG} \\
	& = \SI{170}{\kilo\joule\per\mol} \, . \notag
\end{align}

Note that the production of \ce{NADH} in \Eq{Glu2Pyr} is a side-effect.
This buildup is generally considered the main reason why glycolysis needs to be coupled with additional reactions in order to achieve a proper cyclic process. 
We will discuss small additions to glycolysis in sections~\ref{sec:fermentation} and \ref{sec:gluconeogenesis}, alcoholic fermentation and gluconeogenesis, respectively.

\subsubsection{Tricarboxylic Acid Cycle}
\label{sec:tca-cycle}

\begin{figure*}[tbh]
	\vspace{-1cm}
	  \includegraphics[scale=1]{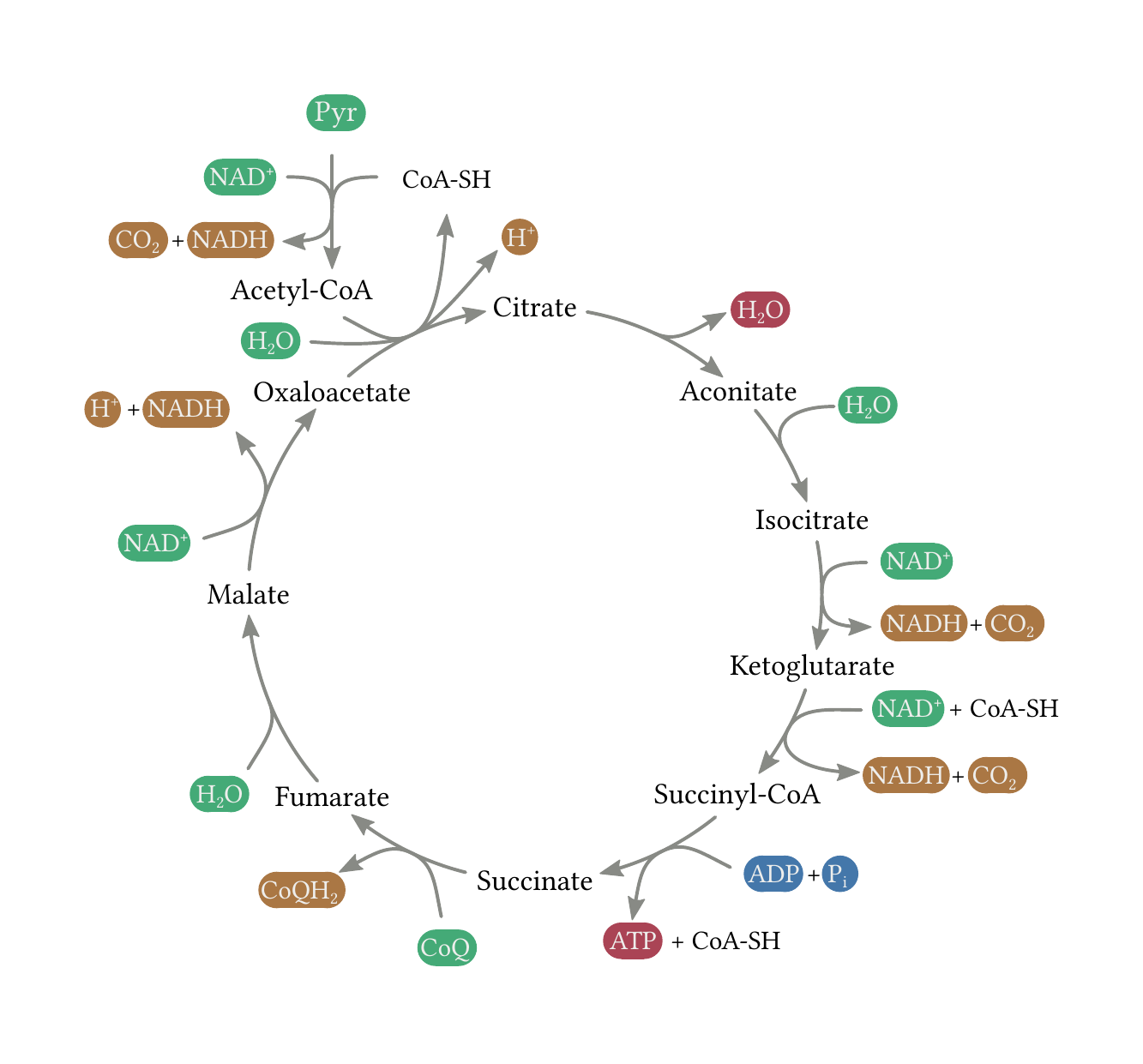}
	\vspace{-1cm}
	\caption{
		Single emergent cycle of the \textbf{tricarboxylic acid (TCA) cycle} which couples the synthesis of \ce{ATP} (red/blue) to the decomposition of Pyruvate (green/brown).
	}
	\label{fig:tca-cycle_forward}
\end{figure*}

The tricarboxylic acid (TCA) cycle is a series of enzymatic reactions \eqref{eq:TCA:reactions} that gradually decompose pyruvate into \ce{CO2} while synthesizing \ce{ATP} and reducing \ce{NAD+} to \ce{NADH}, as well as reducing ubiquinone (\ce{CoQ}) to ubiquinole (\ce{CoQH2}).
The CRN of the TCA cycle is depicted in Fig.~\ref{fig:tca-cycle_forward} together with the single emergent cycle resulting from chemostatting \ce{ Pyr, ATP, ADP, P_{\textrm{i}}, NADH, NAD+, CO2, CoQ, CoQH2, H+}, and \ce{H2O} given by
\begin{gather}
    \ce{ ADP + CoQ + 2 H2O + 4 NAD+ + Pyr + P_{\mathrm{i}} }  \nonumber\\
    \updownharpoons \label{eq:TCA:balance} \\
	\ce{ ATP + 3 CO2 + CoQH2 + 2 H+ + 4 NADH } \nonumber
\end{gather}
This cycle tightly couples two processes:
the synthesis of ATP, \Eq{ATPsynth}, and the anaerobic decomposition of pyruvate
\begin{gather}
  \ce{Pyr + 3 H2O + 4 NAD+ + CoQ} \nonumber \\
  \updownharpoons \label{eq:PyrDec} \\
  \ce{3 CO2 + 4 NADH + CoQH2 + 3 H+} \nonumber
\end{gather}

The free energy released by the decomposition of pyruvate (standard physiological conditions) is about \(\SI{150}{\kilo\joule\per\mole}\) ($60 \, RT$), see \App{CentralEnergyMetabolism}.
Since the whole process is tightly coupled, and just one ATP is hydrolized for any pyruvate decomposed, one finds a modest efficiency:
\begin{equation}
	\eta_{\mathrm{TCA}} = \frac{\Delta_{\mathrm{ATPsynth}}G}{- \Delta_{\mathrm{PyrDec}}G} = \SI{0.30} \, ,
	\label{eq:TCA:eff}
\end{equation}
with
\begin{align}
	\hspace{-1em}
	- \Delta_{\mathrm{PyrDec}}G & = \mu_{\ce{Pyruvate}} + 3\mu_{\ce{H2O}} + \mu_{\ce{CoQ}} + 4\mu_{\ce{NAD+}} \notag \\
	& \; - 3\mu_{\ce{CO2}} - \mu_{\ce{CoQH2}} - 4\mu_{\ce{NADH}} - 3\mu_{\ce{H+}}  \label{eq:PyrDec:DeltaG} \\
	& = \SI{150}{\kilo\joule\per\mol} \, . \notag
\end{align}

\subsubsection{Electron Transport Chain}
\label{sec:electron-transport}

\begin{figure}[tbh]
	\includegraphics[scale=1.0]{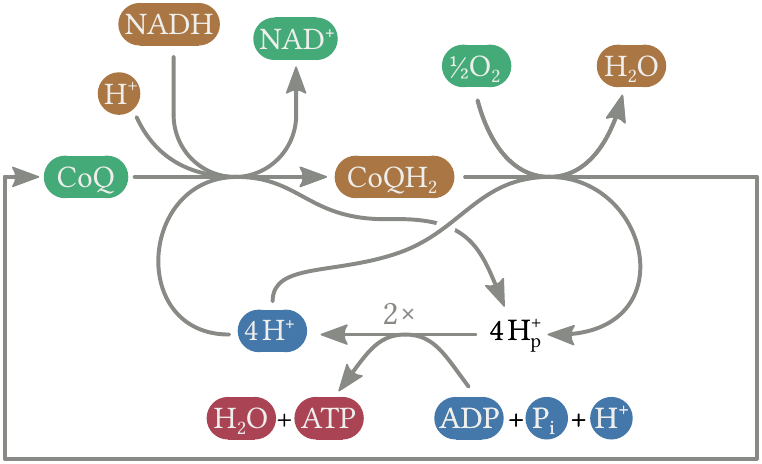}
	\caption{
		 Emergent cycle of the \textbf{electron transport chain} couples the oxidation of \ce{NADH} (green/brown) to the synthesis of ATP (red/blue) via a proton motive force (concentration gradient of protons across the membrane). Note that \ce{CoQ} and \ce{CoQH2} remain exactly balanced in this cycle.
		 For that reason, if they are regarded as internal species, the electron transport chain is characterized solely by this single emergent cycle. Instead, if they are chemostatted, two emergent cycles arise, this one and a second one depicted in Fig. \ref{fig:ETC:forward2}. 
	}
	\label{fig:ETC:forward}
\end{figure}

\begin{figure}[tbh]
	\includegraphics[scale=1.0]{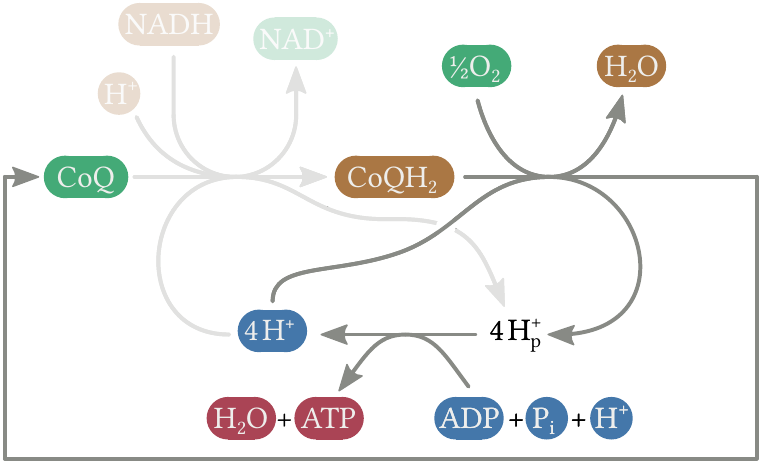}
	\caption{
		With ubiquinol and ubiquinone chemostatted, a second emergent cycle appears in the electron transport chain. It oxidizes ubiquiol (\ce{CoQH2}) and thus restores ubiquinone (\ce{CoQ}) for the TCA cycle.
	}
	\label{fig:ETC:forward2}
\end{figure}

The last part of cellular respiration is the electron transport chain, depicted in Figs.~\ref{fig:ETC:forward} and~\ref{fig:ETC:forward2}, see also \eq{electron-transport-chain--pathway}.
It uses the coenzyme Q in its reduced (\ce{CoQH2}, ubiquinol) and oxidized (\ce{CoQ}, ubiquinone) forms as an intermediate carrier of hydrogen to oxidize \ce{NADH} and thus to restore \ce{NAD+}.
In doing so, the electron transport chain pumps protons from the cytoplasm (\ce{H+}) into the periplasm (\(\ce{H}^{+}_{\mathrm{p}}\)).
The resulting proton motive force is then used by ATP synthase to synthesize ATP from ADP.

We note that, when considering the electron transport chain in isolation, it makes most sense to consider the two forms of coenzyme Q as internal species. In this case the network gives rise to a single emergent cycle (Fig.~\ref{fig:ETC:forward}).
The net effect of this emergent cycle is the synthesis of two \ce{ATP} for each oxidized \ce{NADH}:
\begin{gather}
	\ce{2 ADP + 3 H+ + NADH + \tfrac{1}{2} O2 + 2 P_{\mathrm{i}}} \nonumber\\
	\updownharpoons \label{eq:ETC:balance} \\
	\ce{2 ATP + 3 H2O + NAD+ }\,. \nonumber
\end{gather}
This tightly couples two processes:
the synthesis of \ce{ATP}, \Eq{ATPsynth}, and the oxidation of \ce{NADH},
\begin{align}
	\ce{NADH + \tfrac{1}{2} O2 + H+} &\rightleftharpoons \ce{NAD+ + H2O}\,. \label{eq:NADHoxidat}
\end{align}
The free energy released upon oxidation of \ce{NADH} is about \SI{217}{\kilo\joule\per\mol} (\(88 \, RT\)), and the efficiency is estimated to be 42\%:
\begin{equation}
		\eta_{\mathrm{ETC}} = \frac{2 \Delta_{\mathrm{ATPsynth}}G}{- \Delta_{\mathrm{NADHoxidat}}G}
		= \SI{0.42} \, ,
		\label{eq:ETC:eff}
\end{equation}
with
\begin{align}
	\hspace{-1em}
	- \Delta_{\mathrm{NADHoxidat}}G & = \mu_{\ce{NADH}} + \tfrac{1}{2} \mu_{\ce{O2}} + \mu_{\ce{H+}} - \mu_{\ce{NAD+}} - \mu_{\ce{H2O}} \label{eq:NADHoxidat:DeltaG} \\
	& = \SI{217}{\kilo\joule\per\mol} \, . \notag
\end{align}

The emergent cycle of cellular respiration (\ref{eq:respiration:net-balance}) is composed of reactions from all of the three pathways, and different reactions need to be performed with different multiplicity in order to balance the internal species.
It is however not possible to deduce its net balance solely from the emergent cycles of the three isolated pathways \eqref{eq:glycolysis-net-stoichiometry}, \eqref{eq:TCA:balance}, and \eqref{eq:ETC:balance}.
The reason is that to properly connect the electron transport chain to the TCA cycle it is necessary to also treat both forms of coenzyme Q as chemostatted in the former.
Doing so gives rise to a second emergent cycle in the electron transport chain, depicted in Fig.~\ref{fig:ETC:forward2}, which generates one ATP per \ce{CoQH2}:
\begin{gather}
	\ce{ADP + H+ + CoQH2 + \tfrac{1}{2} O2 + P}_{\mathrm{i}} \nonumber \\
	\updownharpoons \label{eq:ETC:balance2} \\
	\ce{ATP + 2 H2O + CoQ }\,. \nonumber
\end{gather}
As a result, the emergent cycle of cellular respiration (\ref{eq:respiration:net-balance}) is now reproduced by composing:
$1\times$ glycolysis \eqref{eq:glycolysis-net-stoichiometry}, $2\times$ the TCA cycle \eqref{eq:TCA:balance}, $10\times$ the first emergent cycle of the electron transport chain \eqref{eq:ETC:balance}, and $2\times$ the second one \eqref{eq:ETC:balance2}.

\subsection{Ethanol Fermentation}
\label{sec:fermentation}

\begin{figure*}[tbh]
	\vspace{-1cm}
	\includegraphics[scale=1.0]{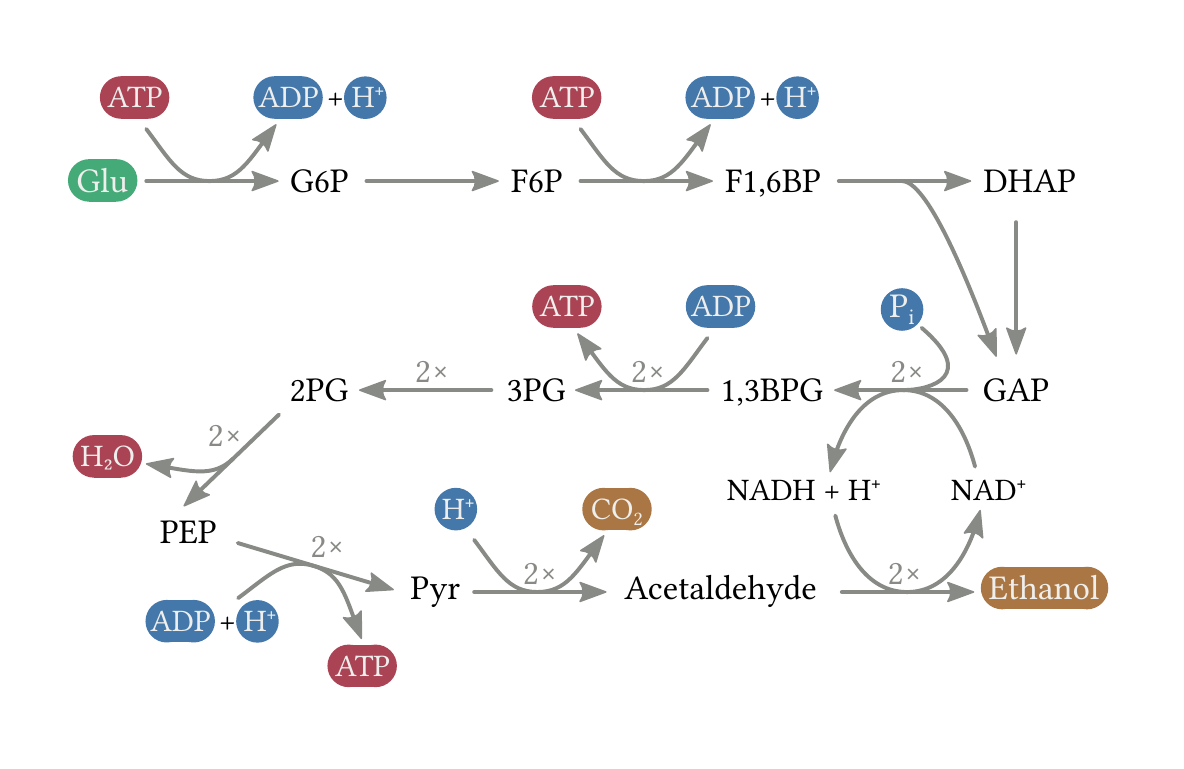}
	\vspace{-1cm}
	\caption{
		Single emergent cycle of \textbf{alcoholic fermentation extending glycolysis} which is the result of a tight coupling of two processes:
		synthesis of \ce{ATP} (red/blue) and decomposition of glucose into ethanol and carbon dioxide (green/brown).
	}
	\label{fig:fermentation-open-forward}
\end{figure*}

Fermentation is a way to re-cycle \ce{NADH} when operating under anaerobic conditions.
This allows the cell to use glycolysis for the synthesis of ATP in a cyclic fashion.
In Fig.~\ref{fig:fermentation-open-forward} we show the reactions of alcoholic fermentation as an emergent cycle.
It shares most of its reactions with regular glycolysis, but extends them by two steps:
cleavage of \ce{CO2} from pyruvate and subsequent addition of hydrogen via oxidation of \ce{NADH}, \eq{ethanolFermentation}.
Thus, when considering the total net balance, ethanol fermentation reads 
\begin{gather}
  \ce{Glu + 2 ADP + 2 P_{\mathrm{i}} + 2 H+}  \nonumber\\
  \updownharpoons \label{eq:EF:balance} \\
  \ce{2 Ethanol + 2 CO2 + 2 ATP + 2 H2O} \nonumber
\end{gather}

We see that \ce{NAD+} and \ce{NADH} need not be regarded as chemostatted:
they are now internal to the reaction network.
In contrast, \ce{CO2} and \ce{Ethanol} take their role as chemostats.
In this form it is easy to see how fermentation tightly couples the two processes
\begin{align}
  \ce{ADP + P_{\mathrm{i}} + H+} &\rightleftharpoons \ce{ATP + H2O} \notag \\
  \ce{Glu} &\rightleftharpoons \ce{2 Ethanol + 2 CO2} \label{eq:Glu2Eth}
\end{align}
with a stoichiometric ratio of 2:1.

Since the conversion of \ce{glucose} into \ce{ethanol} and \ce{CO2}, \Eq{Glu2Eth}, releases roughly \(\SI{267}{\kilo\joule\per\mole}\) ($108 \, RT$), the overall efficiency of such fermentative ATP production is
\begin{equation}
	\eta_{\mathrm{EF}} = \frac{2 \Delta_{\mathrm{ATPsynth}}G}{- \Delta_{\ce{Glu}\rightarrow\ce{Eth}}G}
		= \SI{0.34} \, ,
	\label{eq:EF:eff}
\end{equation}
with
\begin{align}
	\hspace{-1em}
	- \Delta_{\ce{Glu}\rightarrow\ce{Eth}}G & = \mu_{\ce{Glu}} - 2\mu_{\ce{Eth}} - 2\mu_{\ce{CO2}} - 4\mu_{\ce{H+}} \label{eq:Glu2Eth:DeltaG} \\
	& = \SI{267}{\kilo\joule\per\mol} \, . \notag
\end{align}

\subsection{Gluconeogenesis and Futile Cycles}
\label{sec:gluconeogenesis}

\begin{figure*}[h]
	\vspace{-0.5cm}
	\includegraphics[scale=1]{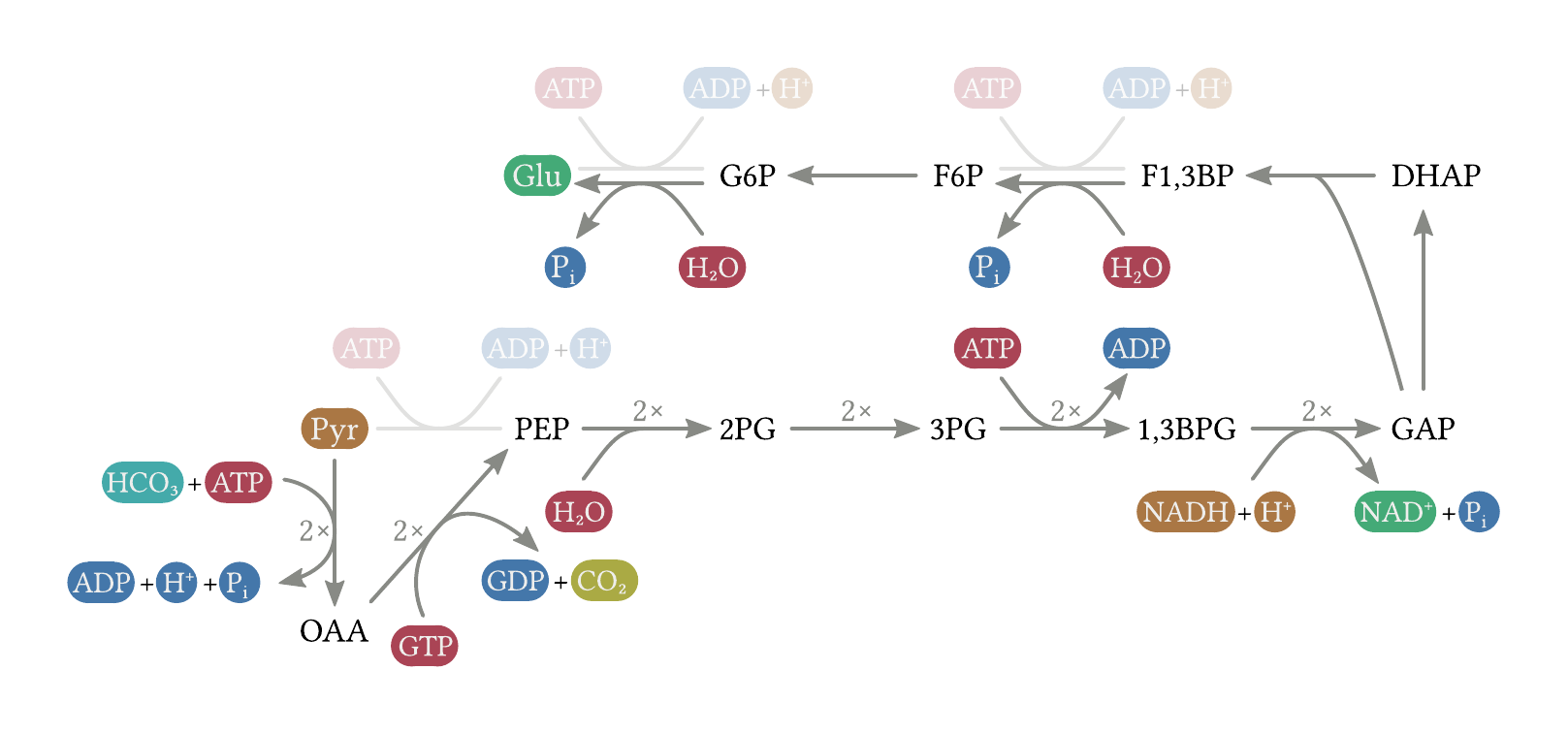}
	\vspace{-0.7cm}
	\caption{
		Emergent cycle of the combined glycolysis and gluconeogenesis network which represents the \textbf{gluconeogenesis pathway}.
		The additional reactions of glycolysis are faded out. 
	}
	\label{fig:gluco}
\end{figure*}

The pathway of gluconeogenesis enables cells to synthesize glucose from pyruvate, see (\ref{eq:glycolysis--reactions}).
The pathway of gluconeogenesis is however not the exact reverse of glycolysis as it involves some different enzymatic reactions and couples to different molecules.
These modifications lead to a different emergent cycle, depicted in Fig.~\ref{fig:gluco}, with net balance
\begin{gather}
	\ce{2 Pyr + 2 NADH + 2 HCO3 + 2 ATP + 2 GTP + 2 H+ + 2 H2O } \nonumber \\
	\updownharpoons \label{eq:gluconeogenesis_net-balance} \\
	\ce{Glu + 2 NAD+ + 2 CO2 + 2 ADP + 2 GDP + 4 P_{\mathrm{i}}} \, . \nonumber
\end{gather}
Gluconeogenesis can be split into two processes: the synthesis of glucose from pyruvate
\begin{align}
	\ce{2 Pyr + 2NADH + 4H+} &\rightleftharpoons \ce{Glu + 2 NAD+} \, ,
	\label{eq:Pyr2Glu}
\end{align}
as well as
\begin{gather}
    \ce{ATP + GTP + HCO3 + H2O} \nonumber\\
    \updownharpoons \label{eq:glucoInput}\\
    \ce{ADP + GDP + 2 P_{\mathrm{i}} + CO2 + H+ } \, . \nonumber
\end{gather}
    Here, \eqref{eq:Pyr2Glu} serves as the output process, while \eqref{eq:glucoInput} is the input process.
    Note that the output process of gluconeogenesis (\textit{i.e.} the synthesis of glucose) is the reverse of the input process of glycolysis \eqref{eq:Glu2Pyr}.
    However, the input process of gluconeogenesis \eqref{eq:glucoInput} involves different molecules than the output process of glycolysis.
Therefore, gluconeogenesis and glycolysis are not exact reversals of each other.

We note that the input process \eqref{eq:glucoInput} could in fact be decomposed further into three separate processes: the hydrolysis of ATP, the hydrolysis of GTP, as well as the dehydration of carbonic acid. But since all three are input processes, separating them all only clutters the notation and adds nothing to the present discussion of gluconeogenesis.

\begin{figure*}[h]
	\vspace{-1cm}
	\includegraphics[scale=1]{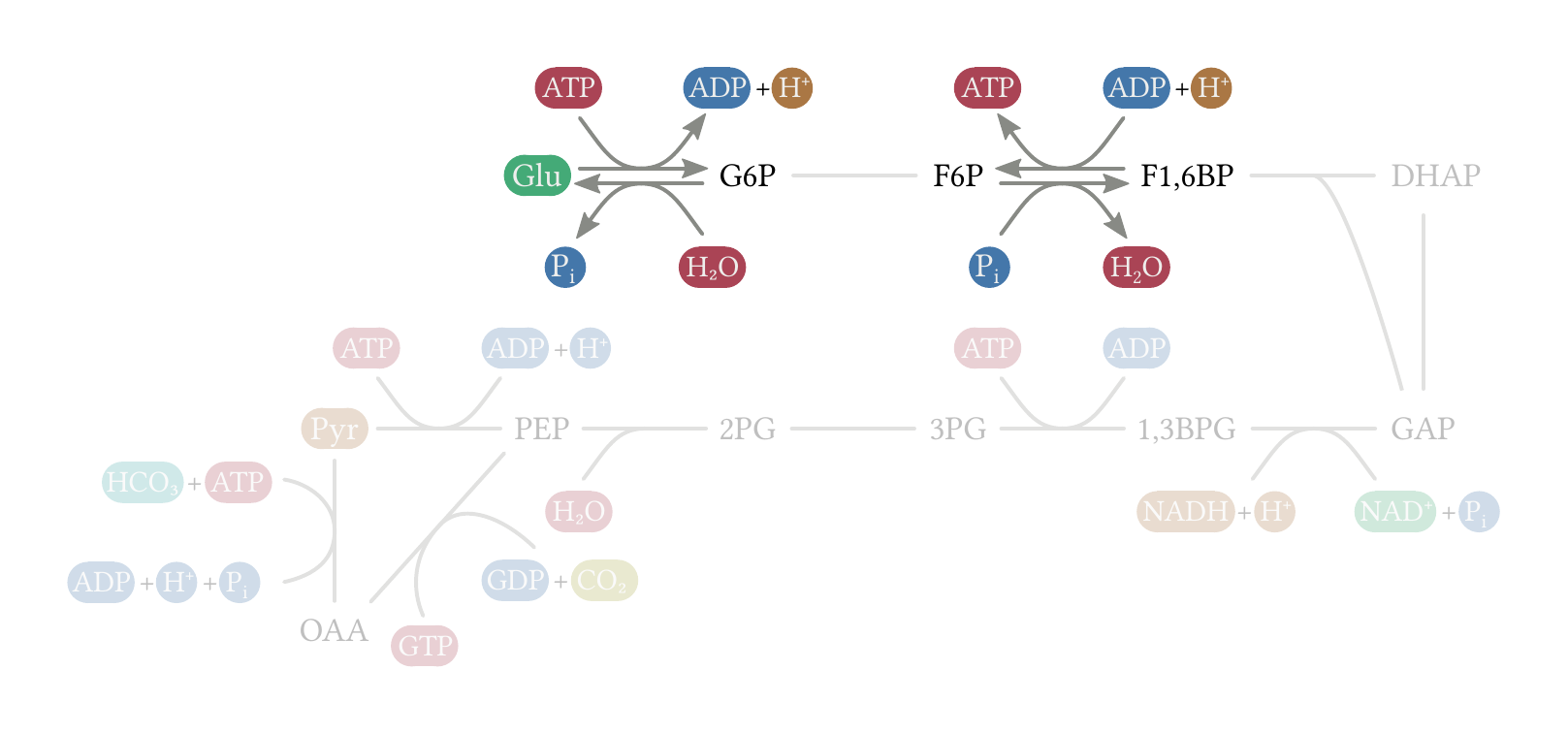}
	\vspace{-0.7cm}
	\caption{
		The internal cycle of the combined glycolysis and gluconeogenesis network.
	}
	\label{fig:gluco:internal}
\end{figure*}

\begin{figure*}[h]
	\vspace{-1cm}
	\includegraphics[scale=1]{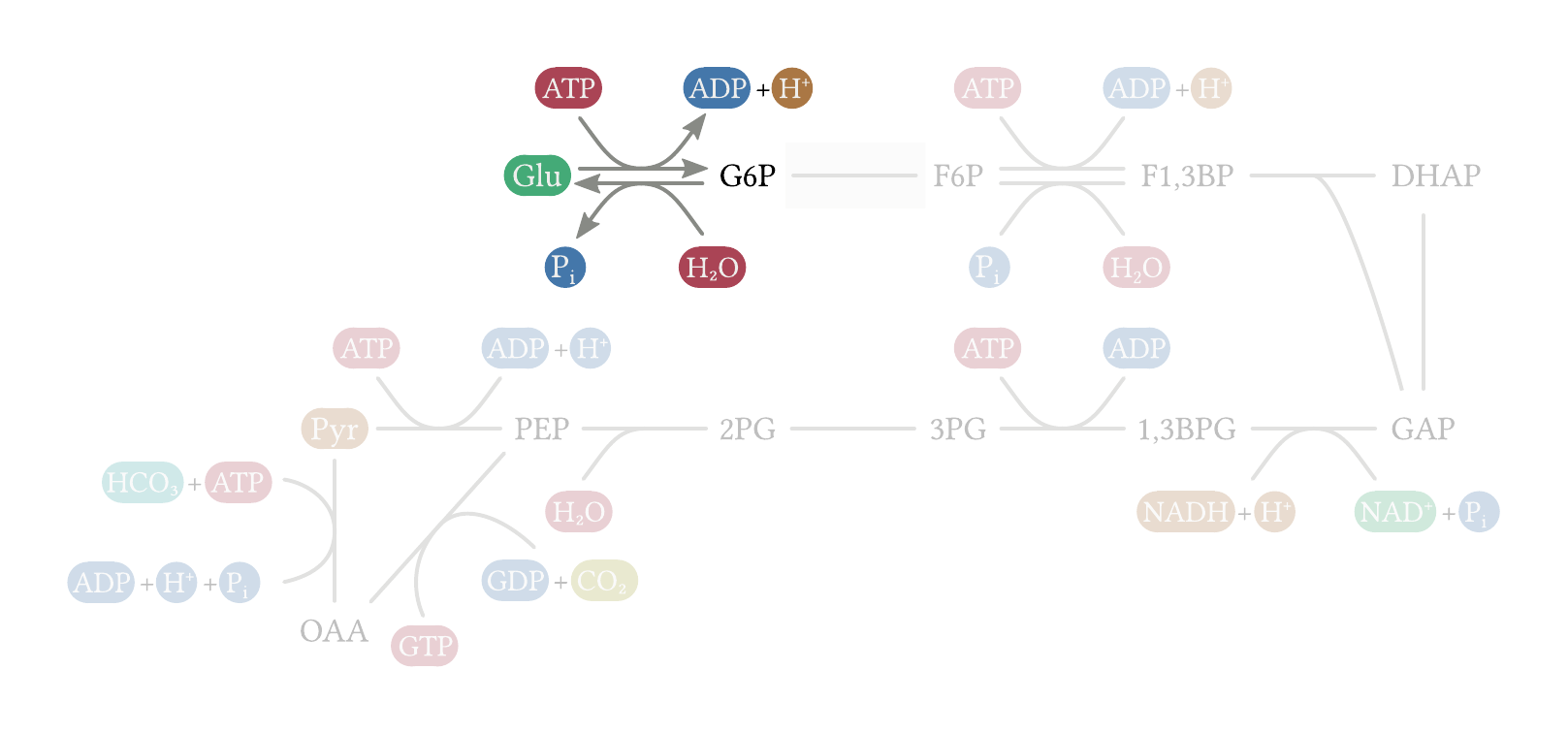}
	\vspace{-0.7cm}
	\caption{
		The emergent cycle of the combined glycolysis and gluconeogenesis network that is commonly referred to as \textbf{futile}.
	}
	\label{fig:gluco:futile}
\end{figure*}

When the CRNs of glycolysis and gluconeogenesis are combined, one obtains three emergent cycles and an internal one.
The three emergent cycles are given by glycolysis (Fig.~\ref{fig:glyco:open:forward}), gluconeogenesis (Fig.~\ref{fig:gluco}), and a \emph{futile cycle} that just hydrolyzes \ce{ATP} to \ce{ADP} without coupling to additional conversions (Fig.~\ref{fig:gluco:futile})
\begin{align}
	\ce{ATP + H2O} &\rightleftharpoons \ce{ADP + P_{\mathrm{i}} + H+} \, .
	\label{futileNet}
\end{align}
The internal cycle (which is also present if the CRN is closed) is depicted in Fig.~\ref{fig:gluco:internal}.
One can easily verify that contrary to emergent cycles, upon completing every reaction in the internal cycle, not only internal species but also chemostatted ones are left unchanged. 

Whether one considers the performance of \ce{ATP} synthesis \eq{ATPsynth} or of glucose synthesis \eq{Pyr2Glu}, tight coupling is broken when the reactions of glycolysis and gluconeogenesis are combined.
Indeed, the dissipation is now carried by three emergent cycles:
\begin{equation}
	T \sigma = - J_{\mathrm{glyco}} \Delta_{\mathrm{glyco}}G - J_{\mathrm{futile}} \Delta_{\mathrm{futile} }G - J_{\mathrm{gluco}} \Delta_{\mathrm{gluco}}G \, ,
\end{equation}
where \(J_{\mathrm{glyco}}\), \(J_{\mathrm{futile}}\), and \(J_{\mathrm{gluco}}\) are the currents associated with each emergent cycle.
The corresponding forces read
\begin{align}
	- \Delta_{\mathrm{glyco}}G 
	    & = - \Delta_{\mathrm{Glu}\rightarrow\mathrm{Pyr}}G - 2 \Delta_{\mathrm{ATPsynth}}G \label{eq:glyco:DeltaG}
	\\
    - \Delta_{\mathrm{futile}}G 
	    & = \Delta_{\mathrm{ATPsynth}}G \label{eq:futile:DeltaG}
	\\
	- \Delta_{\mathrm{gluco}}G
	    & = -\Delta_{\mathrm{Pyr}\rightarrow\mathrm{Glu}}G + 2 \Delta_{\mathrm{ATPsynth}}G \label{eq:gluco:DeltaG}\, ,
\end{align}
with \eqref{eq:Glu2Pyr:DeltaG}, \eqref{eq:ATPsynth:DeltaG}, and
\begin{align}
	- \Delta_{\mathrm{Pyr}\rightarrow\mathrm{Glu}}G
        & =  \Delta_{\mathrm{Glu}\rightarrow\mathrm{Pyr}}G \notag \\
	    & \ \ + 2 (\mu_{\ce{GTP}} + \mu_{\ce{HCO3}} - \mu_{\ce{CO2}} - \mu_{\ce{GDP}} - \mu_{\ce{P_{\mathrm{i}}}})   \, . 
\end{align}
If we now want to characterize the efficiency of ATP synthesis, we need to account for the fact that the output process is now carried by three emergent cycles.
Therefore,
\begin{equation}
	\eta_{\mathrm{ATPsynth}} = \frac{(2 J_{\mathrm{glyco}} - 2 J_{\mathrm{gluco}} - J_{\mathrm{futile}}) \Delta_{\mathrm{ATPsynth}}G}
	{- J_{\mathrm{glyco}} \Delta_{\ce{Glu}\rightarrow\ce{Pyr}}G
	 - J_{\mathrm{gluco}} \Delta_{\ce{Pyr}\rightarrow\ce{Glu}}G
	} \, .
	\label{eq:ATPsynth:eff}
\end{equation}

If $J_{\mathrm{futile}}$ and $J_{\mathrm{gluco}}$ vanish, we recover the tight coupling efficiency of glycolysis alone, \Eq{glyco:eff}.
Under normal physiological conditions, glycolysis and gluconeogenesis are tightly regulated so that they will not occur at the same time \cite{voet16}.
Futile cycles are also known to be regulated. 
Characterizing the effect of these regulations on the efficiency would be very interesting but is beyond the scope of this paper.

We note that most literature on cellular respiration considers the combination of the internal and futile cycle (Figs.~\ref{fig:gluco:internal} and \ref{fig:gluco:futile}) as being two different futile cycles, which both have the net stoichiometry of \ce{ATP} hydrolysis.
Seeing these two cycles as two futile cycles driven by the same force, or as an internal cycle and a futile cycle, is very much a question of taste.
None of the dynamical characterizations will depend on this choice.
Even the total dissipation rate will be invariant under this choice.
However, the values of the individual cycle currents and cycle forces will change.

\section{Defining free-energy transduction and its efficiency}
\label{sec:transduction}

Our work results in the following prescription for the identification of free-energy transduction in CRN described by a stoichiometric matrix \( \stoich \):

\begin{enumerate}
	\item Find the internal cycles in the network---\textit{i.e.\@} combinations of reactions that leave the concentrations of all species unchanged---by finding a basis spanning the kernel of \( \stoich \): \(\set{\vec{c}^{\ell}}\) such that $\stoich \vec{c}^{\ell} = \vec{0}$.
	\item Identify which chemical species are exchanged with the environment---\textit{i.e.\@} the chemostatted species $\ce{Y}$---and split \( \stoich=(\stoich^{\ce{X}}, \stoich^{\ce{Y}})\T \) accordingly:  \(\stoich^{\ce{X}}\) identifies the block corresponding to internal species and \(\stoich^{\ce{Y}}\) to chemostatted species, respectively.
	\item Find the emergent cycles of the network---\textit{i.e.} combinations of reactions that change the concentrations of the chemostatted species but not those of the internal species---by finding the \emph{additional} vectors to \(\set{\vec{c}^{\ell}}\) needed to span the kernel of \(\stoich^{\ce{X}}\): \(\vec{C}\) such that \(\stoich \vec{C} = (\vec{0}, \stoich^{\ce{Y}}\vec{C}) \neq \vec{0}\).
	The net stoichiometries of the emergent cycles are given by \(\stoich^{\ce{Y}}\vec{C}\).
  \item Compute, measure or estimate the currents along the emergent cycles.
  \item Use your knowledge about the charge and atomic composition of the involved molecules to decompose the emergent cycles further:
	  some emergent cycles may be the result of several tightly-coupled processes.
  \item Write the total dissipation as the product of forces and currents of the emergent cycles
    \begin{align}
      T\sigma = - \sum_{\text{cycles}} \Delta\subsc{cyc}G \, J\subsc{cyc} \, ,
    \end{align} 
    and apply the additional decomposition of cycles into processes on the forces:
    \begin{align}
      T\sigma = - \sum_{\text{proc}} \Delta\subsc{proc} G \, J\subsc{proc} \, .
    \end{align}
    In this decomposition all the process currents \(J\subsc{proc}\) that belong to the same emergent cycle are identical to the emergent cycle current \(J\subsc{cyc}\).
  \item Identify the sign of the dissipative contributions \(\set{ - \Delta\subsc{proc} G \, J\subsc{proc} }\).
    Three cases are possible.
    \begin{enumerate}
        \item[(i)] All contributions are positive.
    	    The CRN is purely dissipative.
        \item[(ii)] Some contributions are positive, some are negative.
        The CRN is transducing free energy.
    	The positive contributions are the input, the negative contributions are the output,
    	\begin{align}
            T\sigma = T\sigma\subsc{input} + T\sigma\subsc{output} \geq 0 \, ,
    	\end{align}
    	and the efficiency of free-energy transduction is
    	\begin{align}
            \eta = -\frac{T\sigma\subsc{output}}{T\sigma\subsc{input}} \leq 1 \, .
            \label{eq:definition-efficiency}
    	\end{align}
        \item[(iii)] All contributions vanishing individually is highly unlikely as it would require immense fine tuning of external concentrations to equilibrium values.
    \end{enumerate}
\end{enumerate}

In our considerations we have found one special case for \Eq{definition-efficiency}: Tight coupling.
In this case only one emergent cycle appears, and its current will cancel in the ratio---making the efficiency depend on the ratio of chemical potential differences and thus energetics alone.
In the general case, the efficiency of free-energy transduction depends both on energetics via chemical potentials as well as on the kinetics via the currents of the involved processes.

\section{Discussion and conclusions}
\label{Conc}

We provided the conceptual tools to analyze complex open CRNs as free-energy-transducing chemical machines.
Our approach is based on the knowledge of the stoichiometric matrix and of the species exchanged with the environment (\textit{i.e.} the chemostatted species).
At steady state, only the forces corresponding to the emergent cycles (combinations of chemical potentials of chemostatted species) and their corresponding currents (combinations of currents of chemostatted species) determine the dissipation, see \Eq{EPssEmergent}.
These emergent currents and forces are crucial to define transduction and its efficiency:
they are needed to identify the chemical processes that serve as input and output in the dissipation.

An important outcome of this work is that transduction is generally an emergent phenomenon that occurs at the network level when multiple elementary reactions are coupled to each other.
It typically cannot happen at the level of single elementary reactions, which very often involve mass transfer (see \Sec{massTransfer}).
Two coupled isomerizations may however be seen as an exception to the rule.

We focused our study on steady states, but generalizations to transient and nonautonomous regimes (\textit{e.g.} situations in which the concentrations of the chemostats changes in time) are possible \cite{Rao.Esposito2016,falasco18,avanzini21}.
These generalization have been used for coarse-graining schemes \cite{avanzini20}, as well as in the context of energy storage \cite{penocchio19}.

The framework that we propose in this article provides a general theory of free-energy transduction valid for \emph{nonlinear} CRNs.
Theories of free-energy transduction have been previously established \cite{Hill1977}, but they were limited to linear CRNs.
These theories have been successful in studying molecular motors because the enzymatic dynamics is well described by a linear network in which every transition corresponds to a pseudo-unimolecular reaction \cite{Wachtel.etal2018,Juelicher1997,Parmeggiani1999,Parrondo2002,Seifert2012,Brown2020,amano21}.
By formulating a theory for nonlinear CRN, we paved the way to analyze much more complex CRNs, such as metabolic networks.
These networks contain numerous bi-molecular reactions that could not be described by linear networks.

When applied to metabolic CRNs, our analysis shows that the emergent cycles coincide with known metabolic pathways.
We found that, when taken either separately or together, the three main parts of central energy metabolism---glycolysis, the TCA cycle, and the electron transport chain---correspond to a single emergent cycle, which is distinctive of tightly-coupled free-energy transduction.
We could therefore study their transduction efficiency solely based on stoichiometric and thermodynamic considerations.
However, tight coupling is broken when additional pathways are added---such as gluconeogenesis---as they give rise to additional and possibly futile emergent cycles.
In this case, the transduction efficiencies depend on the currents, and assessing them requires some information about the kinetics. 
This goes beyond the scope of this study but can be envisaged in the future.

Of course, metabolic regulation plays a key role in metabolism and properly describing this aspect remains an open challenge.
Our current thermodynamics framework remains applicable to study any post-translational regulation because such a regulation only affects the kinetics---and not the forces.
In contrast, translational regulations have their own energetics, which would need to be accounted for.
However, the latter operate on longer times scales, and our method could still provide good descriptions for short time scales.

Our method could also be of value for metabolic network reconstruction approaches such as flux or energy balance analysis \cite{beard02,yang05,palsson15}, which seek to determine the reaction currents based on constrained optimizations.
In this context, thermodynamics has been recently recognized as an important factor to constrain reconstruction methods \cite{niebel19,saldida20,oftadeh21,akbari21} but also to study metabolism in general \cite{bar-even12,noor14}.
Further work is required to elucidate these connections.

The largest network we considered in this paper is cellular metabolism.
We now briefly comment on some of the questions that may arise when considering CRNs of larger size and complexity, such as genome-wide metabolic networks or possibly atmospheric and biogeochemical CRNs.
Up to small modifications, the basic principle of our approach should continue to apply.
The question of which species to treat as chemostatted is largely contingent. It is a special instance of the classical question in thermodynamics of what to treat as the system and what to treat as the environment. For example, if the changes in concentration of an internal species in large abundance can be neglected over the relevant time scales of the experiment, one can treat it as chemostatted. The same would be true if external control mechanisms may be at work to control its concentration.
For a given choice of chemostats, the procedure to find the emergent cycles is simple and can be implemented using a simple linear algebra algorithm.
The task is indeed to find a set of independent integer vectors that belong to the null space of the open-system stoichiometric matrix but not to the null space of the closed-system stoichiometric matrix.
As already discussed, this basis is not unique, and some bases may provide clearer insights about how the CRN transduces free energy than other bases.
When composing cellular respiration in terms of glycolysis, TCA cycle and electron transport chain, we have seen that our methodology can be used to study larger CRNs in terms of smaller open CRNs by chemostatting suitable internal species.
This suggest that modular approaches to complex CRN may be developed in the future.

We end by noting that the present work also provides insights on the validity of the second law \eqref{eq:crn:dissipation} and of the local detailed balance \eqref{eq:MAK} for non-elementary reactions.
Indeed, since the latter should be viewed as processes (including emergent cycles), the validity of the second law \eqref{eq:crn:dissipation} is still ensured by \eqref{eq:EPssEmergent}, but not the local detailed balance which imposes the reaction to flow along its force (see Eq. \eqref{eq:MAK} and below).
Instead, as we have seen, processes may undergo transduction and flow against their force if they are coupled to other processes.
To treat a process as if it were an elementary reaction satisfying local detailed balance, one must make sure that when chemostatting its products and reactants, only a single emergent cycle arises which does not share any internal species with any other process in the CRN. This happens for many enzymatic reactions for instance. 
A more systematic discussion on this can be found in Refs.~\cite{Wachtel.etal2018,avanzini20}.  

\section*{Acknowledgements}
 
This work was supported by grants from the Simons Foundation (691552, RR), and the European Research Council, project NanoThermo (ERC-2015-CoG Agreement No.~681456, ME). 

\appendix

\section{Reactions of the Central Energy Metabolism}
\label{sec:CentralEnergyMetabolism}

We expand on the thermodynamic data and the metabolic chemical reactions used in section \ref{sec:metabolism}.

\subsection{Thermodynamics data under physiological conditions}

\begin{table}[t]
  \centering
  \begin{tabular}{l r c l r}
    \toprule
	reactant	& \(\mu\) (\SI{}{\kilo\joule\per\mol})	& \quad   	& reactant	& \(\mu\) (\SI{}{\kilo\joule\per\mol})	\\
    \midrule
	\ce{ATP}	& \(-2281 \pm 3\)		&	& \ce{CO2}		& \(-403 \pm 6\)			\\
	\ce{ADP}	& \(-1406 \pm 2\)		&	& \ce{O2}		& \(-1 \pm 6\)			\\
	\ce{AMP}	& \(-531 \pm 2\)		&	& \ce{CoQ}		& \(790 \pm 20\) 			\\
	\ce{P_{\textrm{i}}}	& \(-1072.6 \pm 1.5 \)	&	& \ce{CoQH2}	& \(770 \pm 20\)  			\\
	\ce{NAD+}	& \(-1146 \pm 13\)		&	& Glucose		& \(-409.4 \pm 1.3\)			\\
	\ce{NADH}	& \(-1080 \pm 13\)		&	& Pyruvate		& \(-355.2 \pm 1.5\)			\\
	\ce{H2O}	& \(-151.9 \pm 1.5 \)	&	& Ethanol		& \(65 \pm 3\)			\\
    \bottomrule
  \end{tabular}
  \caption{
	  Values of chemical potentials used to estimate free energy changes and efficiencies and collected from \emph{eQuilibrator} \cite{noor13:consistent}.
  }
  \label{tab:thermoData}
\end{table}

\begin{table}[t]
  \centering
  \begin{tabular}{l r r }
    \toprule
	Metabolic Process			&\(-\Delta G\) (\SI{}{\kilo\joule\per\mol}) & \(\eta\) \\
    \midrule
	Glucose oxidation			& 2910 \(\pm\) 50 (\(= 1180 \pm 20 \, RT\))		& -- \\
	ATP synthesis				&  -46 \(\pm\) 4 (\(= 18.4 \pm 1.7 \, RT\))		& -- \\
	Cellular respiration		& 1730 \(\pm\) 120 (\(= 700 \pm 50 \, RT\)) 	& 0.41 \(\pm\) 0.04 \\
	Glu\(\rightarrow\)Pyr		&  170 \(\pm\) 40 (\(= 68 \pm 15 \, RT\)) 		& -- \\
	Glycolysis					&   80 \(\pm\) 40 (\(=  31 \pm 15 \, RT\))		& 0.54 \(\pm\) 0.12 \\
	Pyruvate decomposition		&  150 \(\pm\) 80 (\(=  60 \pm 30 \, RT\))		& -- \\
    TCA cycle					&  110 \(\pm\) 80 (\(= 40 \pm 30 \, RT\)) 		& 0.30 \(\pm\) 0.16 \\
	\ce{NADH} oxidation			&  217 \(\pm\) 18 (\(= 88 \pm 8 \, RT\))		& -- \\
	Electron transport chain	&  130 \(\pm\) 20 (\(= 51 \pm 8 \, RT\))		& 0.42 \(\pm\) 0.05 \\
	Glu\(\rightarrow\)Eth		&  267 \(\pm\) 13 (\(= 108 \pm 5 \, RT\)) 		& -- \\
	Ethanol fermentation		&  175 \(\pm\) 16 (\(= 71 \pm 6 \, RT\))		& 0.34 \(\pm\) 0.03 \\
    \bottomrule
  \end{tabular}
	\caption{
		Main metabolic processes discussed in section~\ref{sec:metabolism} and estimates of the corresponding free energy changes under standard physiological conditions.
		For tightly-coupled transduction, we also report the estimated efficiency.
		All errors are evaluated using the standard propagation of errors (square root of the sum of squares).
	}
  \label{tab:eff}
\end{table}

In Tab.~\ref{tab:thermoData}, we report the chemical potentials of the chemostatted species used to estimate thermodynamic forces and efficiencies.
These values capture typical physiological conditions---which we referred to as ``standard physiological conditions''---and are collected from  \emph{eQuilibrator} \cite{noor13:consistent}.
To be specific, the activity coefficient \(\gamma_{i}(\vec{z})\) in the chemical potential for non-ideal dilute solutions, \Eq{mu}, is characterized using the phenomenological \emph{extended Debye--H\"uckel relation}
\begin{equation}
    \ln \gamma_{i}(\vec{z}) = - \frac{a q_{i}^{2} \sqrt{I}}{1 + b \sqrt{I}} \, , \label{eq:EDH}
\end{equation}
where \(q_{i}\) is the charge of the molecule \(i\), \(I\) is the \emph{ionic strength} of the solution, and \(a\) and \(b\) are phenomenological coefficients, see \textit{e.g.} Ref.~\cite[\S 3.6]{Alberty2003}.
Since protons are often regarded as chemostatted---pH fixed---it is convenient to report the values of the chemical potentials using pH as a reference \cite[\S 3.6]{Alberty2003} using the so-called transformed chemical potentials \cite[\S 4.1]{Alberty2003}
\begin{equation}
	\mu_{i}(\vec{z}) = \mu_{i}(\vec{z}) - N_{i} \mu_{\ce{H+}}(\vec{z}) \, ,
\end{equation}
where \(N_{i}\) is the number of proton moieties that the molecule \(i\) contains, and
\begin{equation}
	\mu_{\ce{H+}}(\vec{z}) = \mu^\circ_{\ce{H+}} + RT \mathrm{pH} \ln 10 + RT \ln \gamma_{\ce{H+}}(\vec{z})
\end{equation}
is the chemical potential of free protons written in terms of pH.
A similar discussion holds for magnesium ions \ce{Mg++}, which bind to \(\ce{ATP}\) ions and are also regarded as chemostatted.
The values of chemical potential reported in Tab.~\ref{tab:thermoData} correspond to the choice pH $= 7.5$, pMg $= 3.0$, $I = 0.25$, $z_{i}=10^{-3}$ M (millimolar concentrations, except for water, $z_{\ce{H_2O}}\simeq 55$ M), and $T=298.15$K (\(RT\simeq 2.5\)), see Tab.~\ref{tab:thermoData}.

In Tab.~\ref{tab:eff} we report the overall free energy change, \Eq{crn:force}, and the thermodynamic efficiency, \Eq{tc--efficiency}, for the different tightly-coupled chemical processes discussed in the main text.
%

\subsection{Glycolysis and Gluconeogenesis}

The enzymatic reactions constituting glycolysis are
\begin{subequations}
  \begin{align}
    \ce{           ATP + Glu }&\rightleftharpoons \ce{ ADP + G{6}P + H+ } \\
    \ce{               G{6}P }&\rightleftharpoons \ce{ F{6}P              }        \\
    \ce{         ATP + F{6}P }&\rightleftharpoons \ce{ ADP + F{1,\!6}BP + H+ } \\
    \ce{          F{1,\!6}BP }&\rightleftharpoons \ce{ DHAP + GAP       } \\
    \ce{                DHAP }&\rightleftharpoons \ce{ GAP           }    \\
    \ce{    GAP + NAD+ + P_{\mathrm{i}} }&\rightleftharpoons \ce{ {1,\!3}BPG + H+ + NADH} \\
    \ce{     {1,\!3}BPG + ADP }&\rightleftharpoons \ce{ {3}PG + ATP       }  \\
    \ce{               {3}PG }&\rightleftharpoons \ce{ {2}PG              } \\
    \ce{               {2}PG }&\rightleftharpoons \ce{ H2O + PEP        } \\
    \ce{      ADP + H+ + PEP }&\rightleftharpoons \ce{ ATP + Pyr    }\\
    \intertext{Additionally, gluconeogenesis comprises the reactions}
    \ce{     ATP + HCO3 + Pyr} & \rightleftharpoons \ce{ ADP + H+ + OAA + P_{\mathrm{i}} } \\
    \ce{           GTP + OAA } & \rightleftharpoons \ce{ CO2 + GDP + PEP              } \\
    \ce{    F{1,\!6}BP + H2O } & \rightleftharpoons \ce{ F{6}P + P_{\mathrm{i}}                  } \\
    \ce{         G{6}P + H2O } & \rightleftharpoons \ce{ Glu + P_{\mathrm{i}}       }
  \end{align}
  \label{eq:glycolysis--reactions}
\end{subequations}

\begin{widetext}

When the species \ce{H2O, H+, ATP, ADP, P_{\textrm{i}}, NAD+, NADH, GTP, GDP, CO2, HCO3, Glu, Pyr} are chemostatted, we find three emergent cycles in the glycolysis \& gluconeogenesis network:
\begin{alignat*}{3}
	\text{(A4c)} + \text{(A4m)} &\colon &\ce{ATP + H2O  &\rightleftharpoons  ADP + P_{\mathrm{i}} + H+ } \\
	\text{(A4a--e)} + 2 \: \text{(A4f--j)} &\colon &
	\ce{Glu + 2 ADP + 2 P_{\mathrm{i}} + 2 NAD+}  &\rightleftharpoons \ce{  2 Pyr + 2 ATP + 2 NADH + 2 H+ + 2 H2O } \\
	2 \: \text{(A4k,l)} - 2 \: \text{(A4f--j)} - \text{(A4a--e)} &\colon& & \\
	\ce{2 Pyr + 2 ATP + 2 GTP &+& 2 NADH + 2 HCO3 + 2 H+ + 2 H2O } &\rightleftharpoons \ce{Glu + 2 ADP + 2 CO2 + 2 GDP + 2 NAD+ + 4 P_{\mathrm{i}}}
	\label{eq:glycolysis--cycles}
\end{alignat*}
The first cycle is a \emph{futile cycle} with net balance corresponding to \(\ce{ATP}\) hydrolysis.
The second cycle is the glycolysis pathway decomposing glucose (\(\ce{Glu}\)) into two pyruvate (\(\ce{Pyr}\)) while phosphorylating two \(\ce{ADP}\) into two \(\ce{ATP}\).
The last cycle is the gluconeogenesis pathway producing glucose from pyruvate.
In addition, this system has a purely internal cycle composed of the reactions \(\text{(A4a)}+\text{(A4n)}-\text{(A4m)}-\text{(A4c)}\).

\end{widetext}

\subsection{Tricarboxylic Acid (TCA) Cycle}

The TCA cycle is composed of the Pyruvate dehydrogenase reaction
\begin{subequations}
  \begin{align}
    \ce{          CoASH + NAD+ + Pyr}&\rightleftharpoons \ce{ AcetylCoA + CO2 + NADH  }\\
    \intertext{and the Tricarboxylic Acid Cycle reactions}
    \ce{           AcetylCoA + H2O + OAA }&\rightleftharpoons \ce{ Citrate + CoASH + H+     }\\ 
    \ce{                         Citrate }&\rightleftharpoons \ce{ Aconitate + H2O        }\\   
    \ce{                 Aconitate + H2O }&\rightleftharpoons \ce{ Isocitrate           }\\     
    \ce{               Isocitrate + NAD+ }&\rightleftharpoons \ce{ CO2 + KG + NADH    }\\
    \ce{    CoASH + KG + NAD+ }&\rightleftharpoons \ce{ CO2 + NADH + SucCoA      }\\
    \ce{         ADP + P_{\mathrm{i}} + SucCoA }&\rightleftharpoons \ce{ ATP + CoASH + Succ       }\\
    \ce{                 CoQ + Succ }&\rightleftharpoons \ce{ CoQH2 + Fumarate              }\\
    \ce{                  Fumarate + H2O }&\rightleftharpoons \ce{ Malate                        }\\
    \ce{                   Malate + NAD+ }&\rightleftharpoons \ce{ H+ + NADH + OAA }
  \end{align}
    \label{eq:TCA:reactions}                         
\end{subequations}
When chemostatting \ce{H+}, \ce{ATP}, \ce{ADP}, \ce{P_{\textrm{i}}}, \ce{H2O}, \ce{CO2}, \ce{NAD+}, \ce{NADH}, \ce{CoQ}, \ce{CoQH2}, \ce{Pyr} and when each reaction happens once, we arrive at the single emergent cycle with the net stoichiometry given in \Eq{TCA:balance}.
This cycle represents the oxidation of pyruvate to \ce{CO2} and the simultaneous reduction of ubiquinone (\ce{CoQ}) and \ce{NAD+}.

\subsection{Electron Transport Chain and ATP Synthesis}

The reactions composing the electron transport chain are
\begin{subequations}
  \begin{align}
    \ce{       CoQ + 5 H+ + NADH } &\rightleftharpoons \ce{ CoQH2 + 4 H^+_P + NAD+ }\\
    \ce{     CoQH2 + 4 H+ + {\textstyle\frac{1}{2}} O2 } &\rightleftharpoons \ce{ CoQ + 4 H^+_P + H2O } \\   
    \ce{      ADP + 4 H^+_P + P_{\mathrm{i}} } &\rightleftharpoons \ce{ ATP + 3 H+ + H2O  }
  \end{align}
    \label{eq:electron-transport-chain--pathway}    
\end{subequations}

When chemostatting \ce{ATP}, \ce{ADP}, \ce{P_{\textrm{i}}}, \ce{H+}, \ce{H2O}, \ce{O2}, \ce{NAD+}, \ce{NADH}, \ce{CoQ}, and \ce{CoQH2}, we observe two emergent cycles.
The first originates from the reactions \( \text{(A6a)} + \text{(A6b)} + 2 \: \text{(A6c)} \), and has a net stoichiometry given in \Eq{ETC:balance}.
This cycle corresponds to the oxidation of \ce{NADH} via molecular oxygen in order to synthesize \ce{ATP}.
The second cycle originates from the reactions \( \text{(A6b)} + \text{(A6c)} \), and has a net stoichiometry given in \Eq{ETC:balance2}.
This cycle corresponds to the synthesis of \ce{ATP} via the oxidation of \ce{CoQH2}.
As discussed in the main text, when \ce{CoQ} and \ce{CoQH2} are regarded as internal species, only the first of these two cycles emerges.

\subsection{Ethanol Fermentation}

In ethanol fermentation, the reactions composing glycolysis, Eq.~(\ref{eq:glycolysis--reactions}a--j), are complemented by
\begin{subequations}
  \begin{align}
    \ce{       Pyr + H+ } &\rightleftharpoons \ce{ Ethanal + CO2 }\\
    \ce{     Ethanal + NADH + H+} &\rightleftharpoons \ce{ Ethanol + NAD+ }
  \end{align}
  \label{eq:ethanolFermentation}
\end{subequations}
When replacing \ce{Pyruvate}, \ce{NAD+}, and \ce{NADH} for \ce{Ethanol} as chemostats, one finds the emergent cycle described by the chain of reactions \(\text{(A4a--e)} + 2 \: \text{(A4f--j)} + 2 \: \text{(A7a,b)}\) and whose net balance is given by \Eq{EF:balance}.

\subsection{Cellular Respiration}

Cellular respiration comprise all reactions of glycolysis (without gluconeogenesis) \eq{glycolysis--reactions}, the TCA cycle \eq{TCA:reactions}, and the electron transport chain \eq{electron-transport-chain--pathway}.
By identifying \ce{H2O}, \ce{H+}, \ce{ATP}, \ce{ADP}, \ce{P_{\textrm{i}}}, \ce{Glucose}, \ce{CO2}, \ce{O2} as chemostatted species, cellular respiration can be seen as the chemical process written in \Eq{respiration:net-balance}.
This balance corresponds to the emergent cycle depicted in figure~\ref{fig:energy-metabolism}.
It is composed of the reactions
\text{(A4a--e)} + 2~\text{(A4f--j)}
+ 2~\text{(A5a--j)}
+ 10~\text{(A6a)}
+ 12~\text{(A6b)}
+ 22~\text{(A6c)}\,.

\bibliography{indice}

\end{document}